\documentclass[letterpaper,11pt]{article}
\usepackage{jheppub}
\pdfoutput=1
\usepackage{slashed}
\usepackage{graphicx}
\usepackage{subfigure}
\usepackage{soul}

\usepackage{amsmath}
\usepackage{multirow}
\usepackage{tablefootnote}
\usepackage{hyperref}
\usepackage{epstopdf}
\usepackage{comment}
\usepackage[utf8]{inputenc}

\usepackage[dvipsnames]{xcolor}

\def\figureautorefname~#1\null{Fig.\,#1\null}
\def\tableautorefname~#1\null{Table\,#1\null}
\def\sectionautorefname~#1\null{Section\,#1\null}
\def\subsectionautorefname~#1\null{Section\,#1\null}
\def\equationautorefname~#1\null{Eq.\,(#1)\null}

\def\m1{M_1}
\def\m2{M_2}
\def\m3{M_3}

\def\ch10{\tilde \chi^0_1}

\def\to{\rightarrow}

\newcommand{\lsim}{\mathrel{\mathop{\kern 0pt \rlap
  {\raise.2ex\hbox{$<$}}}
  \lower.9ex\hbox{\kern-.190em $\sim$}}}
\newcommand{\gsim}{\mathrel{\mathop{\kern 0pt \rlap
  {\raise.2ex\hbox{$>$}}}
  \lower.9ex\hbox{\kern-.190em $\sim$}}}

\definecolor{pink}{RGB}{255,105,180}

\def\cosba{\cos(\beta-\alpha)}

\newcommand{\fb}{{\,{\rm fb}}}

\newcommand{\ifb}{\,{\rm fb}^{-1}}

\newcommand{\eehz}{e^+e^- \to hZ}
\newcommand{\eevvh}{e^+e^- \to \nu \bar{\nu} h}

\newcommand{\tanb}{\tan \beta}

\allowdisplaybreaks[4]

\newcommand{\GeV}{\textrm{ GeV}}
\newcommand{\SM}{\textrm{ SM}}

\newcommand{\Br}{\textrm{ Br}}

\RequirePackage[normalem]{ulem}

\title{Comparative Studies of 2HDMs under the Higgs Boson Precision Measurements}

\author[\dagger]{Tao Han,}
\author[\diamond]{Shuailong Li,}
\author[\diamond]{Shufang Su,}
\author[\circ]{Wei Su,}
\author[\#]{Yongcheng Wu}

\affiliation[\dagger]{Department of Physics and Astronomy, University of Pittsburgh,  Pittsburgh, PA 15260, USA}
\affiliation[\diamond]{Department of Physics, University of Arizona, Tucson, Arizona  85721, USA}
\affiliation[\circ]{ARC Centre of Excellence for Particle Physics at the Terascale, Department of Physics,University of Adelaide, South Australia 5005, Australia}
\affiliation[\#]{Ottawa-Carleton Institute for Physics, Carleton University, 1125 Colonel By Drive, Ottawa, Ontario K1S 5B6, Canada}

\preprint{
\begin{flushright}
PITT-PACC-2007   \\
ADP-20-24/T1134
\end{flushright}
}

\emailAdd{than@pitt.edu, shuailongli@email.arizona.edu, shufang@email.arizona.edu, wei.su@adelaide.edu.au, ycwu@physics.carleton.ca}

\abstract {
 We perform comparative studies for four types of the two Higgs Doublet Models (2HDMs) under the precision measurements of the Standard Model (SM) Higgs observables at the proposed   Higgs factories. 
 We explore the discovery potential based on the hypothetical deviations in the precision data for the 2HDMs up to one-loop level. We find  $5\sigma$ observability from the $\chi^2$ fitting in a significant theory parameter space at future Higgs factories.  For the Type-I 2HDM, regions with $\cos(\beta-\alpha)\lesssim -0.1$ or $\cos(\beta-\alpha)\gtrsim 0.08$ are discoverable at more than $5\sigma$ level.  For the other three types of 2HDMs, the $5\sigma$ region is even bigger: $|\cos(\beta-\alpha)|\gtrsim 0.02$ for $\tan\beta\sim 1$.  At small and large values of $\tan\beta$, the region in $\cos(\beta-\alpha)$ is further tightened. We examine the extent to which the different 2HDM theories may be distinguishable from one to the other at the $95\%$ Confidence Level with four benchmark points as case studies.  We  show that a  large part of the parameter space of the other types of 2HDMs can be distinguished from the benchmark points of the target model. The impacts of loop corrections are found to be significant in certain parameter regions. }
\keywords{Higgs precision measurements,  
Beyond the Standard Model, 2HDM.}

\begin{document}
\maketitle
\flushbottom
\clearpage

\section{Introduction}
\label{sec:intro}

With the discovery of the Higgs boson at the CERN Large Hadron Collider (LHC)~\cite{Aad:2012tfa,Chatrchyan:2012xdj}, the Standard Model (SM) of the strong and electroweak (EW) interactions of elementary particles is complete as a self-consistent relativistic quantum field theory potentially valid to exponentially high scales. The agreement between the Standard Model predictions and the experimental observations in particle physics implies that either the new physics beyond the SM is at a higher scale still further from the current experimental reach, or it manifests itself in a more subtle form than those in our simple theoretical incarnations. It is thus prudent to carry out the search both at the energy frontier and at the precision frontier. 

The extension of the SM Higgs sector is theoretically  well-motivated~\cite{Giudice:2008bi,Branco:2011iw}. Exploring the Higgs physics beyond the Standard Model (BSM) is among the high priorities in the current and future experimental programs in high energy physics. Searches for new Higgs bosons at colliders, especially at the LHC~\cite{Aaboud:2017sjh,CMS-PAS-HIG-17-020,Aaboud:2017gsl,Aaboud:2017rel,Sirunyan:2018qlb,Aaboud:2017yyg,Aaboud:2017cxo,Aaboud:2018eoy,Khachatryan:2016are,Aaboud:2018ftw,Sirunyan:2018iwt,ATLAS-CONF-2016-089,Aaboud:2018gjj,CMS-PAS-HIG-16-031,Sirunyan:2019wph,Sirunyan:2019tkw,Aaboud:2019sgt,Sirunyan:2018taj,CMS:2019hvr,Aad:2019zwb}, have been actively conducted, and will remain to be one of the major motivations for future colliders. On the other hand, in the absence of signals of BSM new physics from the current experiments, high precision measurements of the SM parameters, especially the Higgs properties~\cite{Aad:2019mbh,Sirunyan:2018koj}, will sharpen our understanding on physics at the EW scale and provide further insight for new physics.

Recently, there have been lively discussions for construction of Higgs factories to study the Higgs boson properties with high precision. The current proposals include the International Linear Collider (ILC) in Japan \cite{Baer:2013cma,Bambade:2019fyw,Fujii:2019zll,Fujii:2020pxe}, the Circular Electron Positron Collider (CEPC) in China \cite{CEPC-SPPCStudyGroup:2015csa,CEPCStudyGroup:2018ghi} and the electron-positron stage of the Future Circular Collider (FCC-ee) at CERN \cite{Abada:2019lih,Abada:2019zxq,Gomez-Ceballos:2013zzn,fccpara,fccplan,Blondel:2019yqr}. They also have the potential to operate at the $Z$-pole with high luminosities to further improve the existing precision for the SM parameter measurements. With the data samples of a million Higgs bosons and about ${\cal O}(10^{10} - 10^{12})\ Z$-bosons, one would generically expect to achieve a precision for the Higgs property determination of $10^{-3}$, and for the EW observables of $10^{-6}$, deeply into the quantum and virtual contributions from possible new physics effects.

There have been many studies in the literature on the implications of the Higgs precision measurements at current and future colliders on the 2HDMs~\cite{Barger:2014qva,Johansen:2015nxa,Han:2017pfo,Kanemura:2018yai,Kanemura:2019kjg,Chun:2019sjo,Kling:2020hmi,Arco:2020ucn}.  In recent works~\cite{Gu:2017ckc,Chen:2019pkq, Chen:2018shg,Su:2019ibd,Kanemura:2015mxa,Braathen:2019zoh}, we examined the achievable sensitivity of the Higgs and $Z$ factories to probe the virtual effects of the two Higgs doublet model (2HDM) of Type-II and Type-I. With  multivariable $\chi^2$-fit,  we found interesting results in setting significant bounds in a large theory parameter space beyond the reach of the high luminosity upgrade of the LHC (HL-LHC). In this paper, we take the analyses to the next stage. We examine four types of the 2HDMs, namely, Type-I, Type-II, Type-L (lepton specific) and Type-F (flipped Yukawa couplings). 
We establish the theory parameter regions for $5\sigma$ discovery from the deviations from the SM expectations based on the expected precision at future Higgs factories, including the one-loop effects.  Once achieving the signal observation in certain favorable parameter region, we explore the ability to distinguish different types of the  2HDMs with respect to the observables of Higgs  precision measurements.   
It is quite informative that the characteristic features of each 2HDM, primarily their Yukawa couplings, would be notably reflected by the corresponding observables.  

The rest of the paper is organized as follows. In \autoref{sec:model}, we introduce the theoretical framework of the four types of the 2HDMs. In \autoref{sec:study},  we review the current Higgs measurements and future precision expectations adopted in our analyses.  We then present our fitting methodology. Going beyond the existing studies in the literature, we present the current LHC 95\% Confidence Level (C.L.) allowed region as well as the  $5\sigma$ discovery potential at future Higgs factories for the four types of the 2HDMs in \autoref{sec:discover}.  Because of the qualitative differences among the 2HDMs considered here, we demonstrate in \autoref{sec:impl} the feasibility to distinguish the theoretical models from each other based on the precision measurements of different observables.  We summarize the results and draw our conclusions in \autoref{sec:conclude}.
 
\section{Two Higgs Doublet Models}
\label{sec:model}

The Higgs sector of the 2HDMs \cite{Branco:2011iw} consists of two SU(2)$_L$ scalar doublets  $\Phi_i\,(i=1,2)$ with hyper-charge $Y=1/2$
\begin{align}
        \Phi_i = \left(\begin{array}{c}
                \phi_i^+\\
                (v_i+\phi_i^0+iG_i^0)/\sqrt{2}
        \end{array}\right)
\end{align}
where $v_i\,(i=1,2)$ are the vacuum expectation values (vev) of the doublets after the electroweak symmetry breaking (EWSB), satisfying $\sqrt{v_1^2+v_2^2}=v=246 \GeV$.

The 2HDM Lagrangian for the Higgs sector is given by
\begin{align}
        \mathcal{L} = \sum_i|D_\mu\Phi_i|^2 - V(\Phi_1,\Phi_2) + \mathcal{L}_{\rm Yuk}
\end{align}
where $D_\mu$ is the covariant derivative, $V(\Phi_1,\Phi_2)$ is the scalar potential, and $\mathcal{L}_{\rm Yuk}$ contains the Yukawa couplings.

The most general CP-conserving potential with a soft $\mathbb{Z}_2$ symmetry breaking term ($m_{12}^2$) is
\begin{align}
        V(\Phi_1,\Phi_2) =& m_{11}^2 \Phi_1^\dagger\Phi_1 + m_{22}^2\Phi_2^\dagger\Phi_2 - (m_{12}^2\Phi_1^\dagger\Phi_2 + h.c.) + \frac{\lambda_1}{2}(\Phi_1^\dagger\Phi_1)^2 + \frac{\lambda_2}{2}(\Phi_2^\dagger\Phi_2)^2 \nonumber \\
        & + \lambda_3(\Phi_1^\dagger\Phi_1)(\Phi_2^\dagger\Phi_2) + \lambda_4(\Phi_1^\dagger\Phi_2)(\Phi_2^\dagger\Phi_1) + \left[\frac{\lambda_5}{2}(\Phi_1^\dagger\Phi_2)^2 + h.c.\right] .
\end{align}
After the EWSB, the scalars mix with each other to form the mass eigenstates:
\begin{align}
    \left(\begin{array}{c}
        H^\pm\\
        G^\pm 
    \end{array}\right) = \left(\begin{array}{cc}
        c_\beta & -s_\beta \\
        s_\beta & c_\beta
    \end{array}\right)\left(\begin{array}{c}
        \phi_2^\pm\\
        \phi_1^\pm
    \end{array}\right),\\
    \left(\begin{array}{c}
        A^0\\
        G^0 
    \end{array}\right) = \left(\begin{array}{cc}
        c_\beta & -s_\beta \\
        s_\beta & c_\beta
    \end{array}\right)\left(\begin{array}{c}
        G_2^0\\
        G_1^0
    \end{array}\right),\\
    \left(\begin{array}{c}
        h\\
        H 
    \end{array}\right) = \left(\begin{array}{cc}
        c_\alpha & -s_\alpha \\
        s_\alpha & c_\alpha
    \end{array}\right)\left(\begin{array}{c}
        \phi_2^0\\
        \phi_1^0
    \end{array}\right),
\end{align}
where $\alpha$ and $\beta$ are the mixing angles, and $\tan\beta = v_2/v_1$ at tree level. Instead of the eight parameters in the scalar potential $m_{11}^2,m_{22}^2,m_{12}^2,\lambda_{1,2,3,4,5}$, a more convenient set of the parameters is $v,\tan\beta,\cosba,m_h,m_H,m_A,m_{H^\pm}, m_{12}^2$, where $m_h,m_H,m_A,m_{H^\pm}$ are the physical masses of the corresponding Higgs bosons. The relations between these two sets of parameters can be found in Ref.~\cite{Gunion:2002zf}.

\begin{table}[!tbp]
    \begin{center}
        \begin{tabular}{ccccccccc}
            \hline\hline                    
            Types   & $\Phi_1$ & $\Phi_2$ & $u_R$ & $d_R$ & $\ell_R$ & $Q_L$, $L_L$&$\Phi_1$ & $\Phi_2$ \\
            \hline
            Type-I & $+$ & $-$ & $-$ & $-$ & $-$ & $+$&&$u$, $d$, $\ell$ \\
            Type-II & $+$ & $-$ & $-$ & $+$ & $+$ & $+$ &$d$, $\ell$& $u$\\
            Type-L & $+$ & $-$ & $-$ & $-$ & $+$ & $+$ & $\ell$&$u$, $d$,\\ 
            Type-F & $+$ & $-$ & $-$ & $+$ & $-$ & $+$ &$d$&$u$, $\ell$\\
            \hline\hline
        \end{tabular}
        \caption{Four types of  assignments for the $\mathbb{Z}_2$ charges and Yukawa couplings for the scalar doublets $\Phi_{1,2}$ and the SM fermions.}
        \label{tab:z2_assign}
    \end{center}
\end{table}

\begin{table}[!tbp]
\begin{center}
\resizebox{0.6\textwidth}{!}{
{\renewcommand\arraystretch{1.2}
\begin{tabular}{c|cccc}\hline\hline
&\multicolumn{4}{c}{Tree-level Normalized Higgs couplings}\\\cline{2-5}
&$\kappa_h^u$&$\kappa_h^d$&$\kappa_h^e$&$\kappa_h^V$\\\hline
Type-I&$\frac{\cos\alpha}{\sin\beta}$&$\frac{\cos\alpha}{\sin\beta}$&$\frac{\cos\alpha}{\sin\beta}$&$\sin(\beta-\alpha)$\\\hline
Type-II&$\frac{\cos\alpha}{\sin\beta}$&$-\frac{\sin\alpha}{\cos\beta}$&$-\frac{\sin\alpha}{\cos\beta}$&$\sin(\beta-\alpha)$\\\hline
Type-L&$\frac{\cos\alpha}{\sin\beta}$&$\frac{\cos\alpha}{\sin\beta}$&$-\frac{\sin\alpha}{\cos\beta}$&$\sin(\beta-\alpha)$\\\hline
Type-F&$\frac{\cos\alpha}{\sin\beta}$&$-\frac{\sin\alpha}{\cos\beta}$&$\frac{\cos\alpha}{\sin\beta}$&$\sin(\beta-\alpha)$\\\hline\hline
\end{tabular}}}
\caption{Higgs couplings to the SM fermions in the four types of 2HDMs, normalized to the corresponding SM values~\cite{Branco:2011iw}. 
}
\label{tab:kappas}
\end{center}
\end{table}

The Yukawa couplings of the two Higgs doublets are given by 

\begin{equation}
-\mathcal{L}_{Y}=Y_{u} \bar{Q}_{L}\tilde{\Phi}_{u} u_{R}+Y_{d} \bar{Q}_{L} \Phi_{d} d_{R}+Y_{e} \bar{L}_{L} \Phi_{e} e_{R}+\mathrm{h.c.}, 
\end{equation}
with $\tilde{\Phi}=i\sigma_2\Phi^*$ and $\Phi_{u,d,e}$ are either $\Phi_{1}$ or $\Phi_{2}$.
To avoid tree-level  flavor-changing-neutral-currents (FCNCs), a discrete $\mathbb{Z}_2$ symmetry\footnote{This symmetry is broken by the $m_{12}^2$  term in the scalar potential.}  is imposed. There are four possible choices for the charge assignment of the fermions under $\mathbb{Z}_2$, which are shown in \autoref{tab:z2_assign}, along with the non-zero Yukawa couplings for each $\Phi$.  

Expanding the 2HDM Lagrangian after EWSB and rotating into mass eigenstates, we have, for the gauge and Yukawa couplings of the SM-like Higgs boson\footnote{In our analyses, we take the light CP-even Higgs as the 125 GeV SM-like Higgs boson.} $h$, 
\begin{align}
    \mathcal{L} = \kappa_Z \frac{m_Z^2}{v}h Z_\mu Z^\mu + \kappa_W \frac{2m_W^2}{v}h W^+_\mu W^{\mu -} - \sum_{f=u,d,\ell}\kappa_f h \bar{f}f.
\end{align} 
At tree level, $\kappa_i=\kappa_h^i$ only depend on two mixing angles ($\alpha$, $\beta$) and are listed in \autoref{tab:kappas}.
Note that the two normalized Yukawa couplings can be instead expressed in terms of two more commonly used parameters $\cos(\beta-\alpha)$ and $\tan\beta$ as 
\begin{align}
\frac{\cos \alpha}{\sin\beta}&=\sin (\beta-\alpha)+\cot\beta \cos (\beta-\alpha), \nonumber\\
-\frac{\sin \alpha}{\cos\beta}&=\sin (\beta-\alpha)-\tan\beta \cos (\beta-\alpha).
\end{align}

With the anticipated high precision of Higgs coupling measurements at future Higgs factories, they are sensitive to radiation corrections. The one-loop corrections are calculated for the Higgs couplings with the on-shell scheme~\cite{FeynArts-SM,Kanemura:2004mg,Kanemura:2015mxa} using {\tt FeynArts}~\cite{Hahn:2000kx}, {\tt FormCalc}~\cite{Hahn:2016ebn}, {\tt FeynCalc}~\cite{Shtabovenko:2016sxi,Mertig:1990an} and {\tt LoopTools}~\cite{Hahn:1998yk} which are cross-checked with {\tt H-COUP}~\cite{Kanemura:2017gbi} and with {\tt 2HDECAY}~\cite{Krause:2018wmo}.
All the couplings ($\kappa_i$'s) at loop level depend on the mass parameters of the other heavy states running in the loops, as well as the soft $\mathbb{Z}_2$  breaking parameter $m_{12}^2$,  in addition to the parameters $\alpha$ and $\beta$.

\section{Study Strategy}
\label{sec:study}

\subsection{Precision measurements at future colliders}
 
The properties of the SM-like Higgs boson are measured at the current LHC Run-II~\cite{Aad:2019mbh, Sirunyan:2018koj}, and will be measured to a high precision at future Higgs factories~\cite{Cepeda:2019klc,deBlas:2019rxi}.
In the previous works \cite{{Gu:2017ckc,Chen:2019pkq, Chen:2018shg,Su:2019ibd}},  we studied the implications of the anticipated precision measurements on various new physics models, such as the singlet extension of the SM, 2HDM, and composite Higgs model, assuming that no deviation is observed at future Higgs factories.  Naturally, we would like to explore the discovery potential of the Higgs factories for BSM physics.
We will take the estimated precisions for Higgs coupling measurements at  the HL-LHC with {3} ab$^{-1}$ integrated luminosity each from the ATLAS and CMS measurements~\cite{Cepeda:2019klc} and CEPC program with 5.6 ab$^{-1}$   integrated luminosity~\cite{Chen:2019pkq,CEPCStudyGroup:2018ghi,CEPCPhysics-DetectorStudyGroup:2019wir}.    Generally speaking, three future Higgs factories (CEPC, FCC-ee, and ILC) have compatible precisions for Higgs property  measurements, especially for the cross section of $\eehz$ and the signal strength of $\eehz$ with $h\to bb$. While Higgs production at the CEPC~\cite{CEPC-SPPCStudyGroup:2015csa,CEPCStudyGroup:2018ghi} and the FCC-ee~\cite{Abada:2019lih,Abada:2019zxq,Gomez-Ceballos:2013zzn,fccpara,fccplan,Blondel:2019yqr} is dominated by $\eehz$ near  240 GeV$-$250 GeV,  FCC-ee, as well as ILC~\cite{Baer:2013cma,Bambade:2019fyw,Fujii:2019zll,Fujii:2020pxe}, may accumulate more data with $\eevvh$ via $WW$ fusion at higher energies.
Furthermore, ILC running at higher energy may have the access to the self-coupling $\lambda_{hhh}$~\cite{Maltoni:2018ttu}. A summary of the latest Higgs precision measurements of signal strength at future Higgs factories that is used in our analyses can be found in Ref.~\cite{Chen:2019pkq,deBlas:2019rxi}.  
 
\begin{table}[!t]
\centering
\begin{tabular}{l | c c c c}
\hline
\hline 
Collier &LHC Run-II &LHC& HL-LHC & CEPC\\
\hline
$\sqrt{s}$ &13 TeV & 14 TeV &14 TeV& 240 GeV \\
$\int\mathcal{L}dt$ & 80 fb$^{-1}$ & 300 fb$^{-1}$&3 ab$^{-1}$& 5.6 ab$^{-1}$\\
\hline
$\kappa_\tau$&16\%&14\%&1.9\%&1.3\%\\
\hline
$\kappa_b$&19\%&23\%&3.7\%&1.2\%\\
\hline
$\kappa_t$&15\%&22\%&3.4\%&- \\
\hline
$\kappa_c$&-&-&-&2.1\%\\
\hline
$\kappa_W$&9\%&9.0\%&1.7\%&1.3\%\\
\hline
$\kappa_Z$&8\%&8.1\%&1.5\%&0.13\%\\
\hline
$\kappa_g$&11\%&14\%&2.5\%&1.5\%\\
\hline
$\kappa_\gamma$&9\%&9.3\%&1.8\%&3.7\%\\
\hline
\hline
\end{tabular} 
\caption{Estimated statistical precision on normalized Higgs couplings at current LHC Run-II in ATLAS~\cite{Aad:2019mbh}, LHC with 300 $\fb^{-1}$ integrated luminosity at $\sqrt{s}=14$ TeV \cite{ATL-PHYS-PUB-2014-016}, HL-LHC with integrated luminosity of 3 ab$^{-1}$ at 14 TeV (including  both ATLAS and CMS) \cite{Cepeda:2019klc} and CEPC with integrated luminosity of 5.6 ab$^{-1}$ at 240 GeV \cite{CEPCStudyGroup:2018ghi}.  }
\label{tb:exp_kappa}
\end{table}

The estimated precision on normalized Higgs couplings at the current LHC Run-II at ATLAS~\cite{Aad:2019mbh}, LHC with 300 $\fb^{-1}$ integrated luminosity at $\sqrt{s}=14$ TeV \cite{ATL-PHYS-PUB-2014-016}, HL-LHC~\cite{Cepeda:2019klc} and CEPC~\cite{CEPCStudyGroup:2018ghi} are listed in \autoref{tb:exp_kappa}~\footnote{Note that the precisions for the  latest LHC Run-II results of $\kappa_{b, t, g}$  are better than the available predictions at the LHC $300\ifb$ from 2014 analyses.}.  For the up-type Yukawa couplings, $\kappa_t$ is measured at the LHC while $\kappa_c$ will be measured at CEPC given the lower center of mass energy  and clean experimental environment at a lepton collider. While all the couplings can only be determined at a 8\%$-$20\% level at the LHC, the HL-LHC can improve the precision significantly to a few percent level.  CEPC has the best precision for $\kappa_Z$ at sub-percent level, and about 1\% for $\kappa_{\tau, b,W, g}$.  The precision for $\kappa_c$ and $\kappa_\gamma$ is a bit worse given the limited statistics.

\subsection{Fitting method} 
While the previous works focus on the scenarios when no deviation from the SM predictions is observed at future Higgs factories \cite{Gu:2017ckc,Chen:2019pkq,Chen:2018shg,deBlas:2019wgy,An:2018dwb,DiVita:2017vrr,Liebler:2016ceh}, in this work, we examine the extent to which deviations from the SM predictions can be observed, reaching 5$\sigma$ discovery sensitivity for the 2HDMs. We further explore how different types of 2HDMs can be distinguished after the $5\sigma$ observation, as well as how to narrow down the parameter spaces for a given type of 2HDM.

To perform a $\chi^2$-fit,   we adopt the  signal strength modifier (SSM) $\mu$: 
\begin{equation}
\mu=\frac{\sigma\times \Br}{(\sigma\times \Br)_\SM}
\label{eq:SSM}
\end{equation}
of the SM-like Higgs boson in different production and decay channels to parameterize the prediction of different models as well as the experimental data. As a test statistic, we use
\begin{equation}
\chi^2=\sum_i\frac{(\mu_i^0-\mu_i^1)^2}{\sigma_{\mu_i}^2}, 
\label{eq:chi2}
\end{equation}
for $\mu_i^0$ being the prediction of $\mu_i$ in a given testing model,  $\mu_i^1$ being the experimentally observed value, and $\sigma_{\mu_i}$ being the corresponding experimental precision.  For future colliders, $\mu_i^1$ is taken to be the SM values 1, if assuming no deviation from the SM value is observed, or a specific set of values when certain deviations are assumed.  In the case when the experimental observed data is taken to be the prediction of another model, $\chi^2$ could be interpreted as the capability to distinguish them.  We assume the $\chi^2$ statistic follows the $\chi^2$ probability distribution function (p.d.f) with the number of degrees of freedom (d.o.f.) equal to the number of fitting observables $\mu$ subtracted by the number of internal variables of the testing model.   As such, a 5$\sigma$ discovery or a 95\% C.L. exclusion corresponds to a (two-tail) $p$-value of $5.7\times 10^{-7}$ or 0.05, respectively. 

For the current LHC measurements, we use $\Delta\chi^2=\chi^2-\chi_{\min}^2$  to obtain the 95\% C.L. exclusion, in which $\chi_{\min}^2$ is the corresponding $\chi^2$ of the best fitting point of the testing model to the experimentally observed value.  In this case, the number of d.o.f of the $\chi^2$ statistic  is equal to the number of internal variables of the testing model. 

\section{Discovery Potential and Characteristics of 2HDMs}
\label{sec:discover}

In this section we first present  the 95\% C.L. allowed regions in the parameter space of 2HDMs given the current LHC limit, and then explore the 5$\sigma$ discovery regions at the $300\fb^{-1}$ LHC,  the HL-LHC,  and Higgs factories such as the CEPC. In the due course, we discuss the unique characteristics of different 2HDMs.
 
\begin{figure}[!tbp]
\centering
\includegraphics[width=0.49\textwidth]{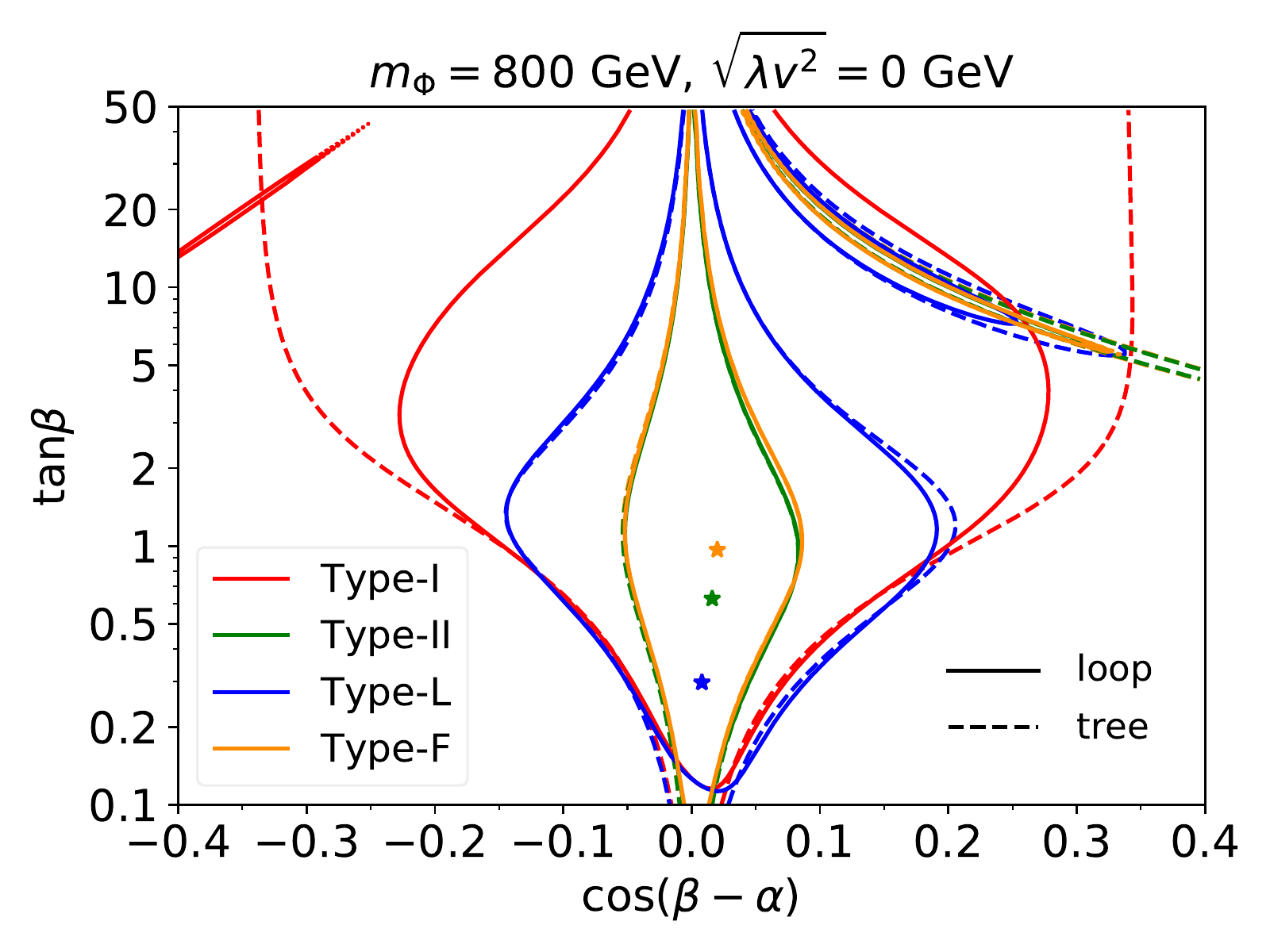}
\includegraphics[width=0.49\textwidth]{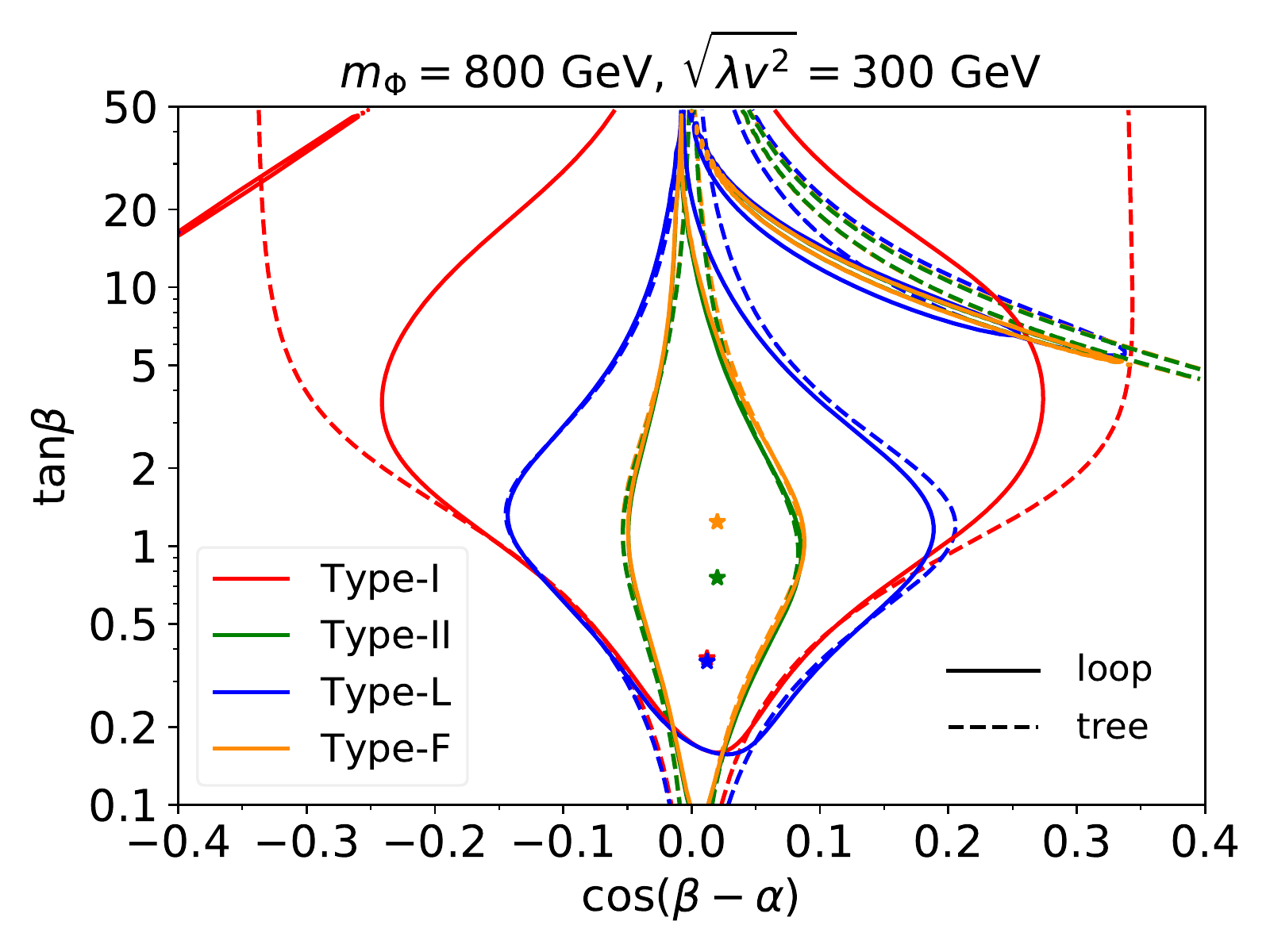}
\caption{95\% C.L. allowed regions enclosed by the solid (dashed) curves from one-loop (tree level) results under the current LHC limits. Results of Type-I, II, L and F are indicated by the red, green, blue and orange colors, respectively, with the one-loop level best fitting point of each type marked by a star of the corresponding color. For loop results, we assume a degenerate non-SM Higgs mass $m_\Phi=m_H=m_A=m_{H^\pm}=800$ GeV, with $\sqrt{\lambda v^2}=0$ in the left panel and 300 GeV in the right panel.  }
\label{fig:lhcrun2limit}
\end{figure}

 The allowed 95\% C.L. regions for four types of 2HDMs under the current LHC limit \cite{Aad:2019mbh} in $\cos(\beta-\alpha)$-$\tan\beta$  are shown in \autoref{fig:lhcrun2limit}, using the $\Delta\chi^2$ statistic  with the number of d.o.f$=2$ for the two fitting parameters $\tan\beta$ and $\cos(\beta-\alpha)$.  The results of Type-I, II, L and F are indicated by red, green, blue and orange colors, respectively.   The regions enclosed by the solid (dashed) curves are the 95\% C.L. allowed regions at one-loop (tree) level, with the one-loop level best fitting point   of each type marked by a star of the corresponding color. For the loop-level results, we assume a degenerate non-SM Higgs mass $m_\Phi=m_H=m_A=m_{H^\pm}=800$ GeV, with $\sqrt{\lambda v^2}\equiv \sqrt{m_H^2-m_{12}^2/(\sin\beta\cos\beta)}=0$ in the left panel and 300 GeV in the right panel.

The behavior of Type-II resembles that of Type-F,  while Type-I resembles Type-L except that the allowed region of Type-I opens up for large $\tan\beta$.  This is because the difference between the two corresponding 2HDM types is in the lepton sector, and the impact of the leptonic couplings is small, given the dominating bottom decay branching fractions, and the comparable measured precision of $\tau\tau$ channels with $bb$ channel at the LHC. 

Those band-shaped regions in the upper right quadrant are the  ``wrong-sign" regions~\cite{Ferreira:2014naa}. This is in the neighbourhood of $\cosba\sim {2}/{\tanb}$, where the Yukawa couplings induced by $\Phi_2$ in \autoref{tab:kappas}
happen to be unity, and those  induced by $\Phi_1$ are near $-1$, thus the name of ``wrong-sign". 
$\Delta\kappa_i$ is small and changes sign when passing it from the left to the right for $\Phi_2$ induced Yukawa couplings.  The wrong-sign region of Type-I is hidden inside the wide allowed region at large $\tan\beta$.  Again, we observe the similarity between Type-II and Type-F given the same quark coupling structure, which results in the same loop-induced $hgg$ corrections as well.

There is another wrong-sign region along the line $\cosba=-2\tanb$ in the small $\tan\beta$ region, where the Yukawa couplings induced by $\Phi_2$ are $-1$, while $\Phi_1$ induced couplings are 1, instead.  Such regions,  however, are severely constrained by the large corrections to the loop-induced $\kappa_{h\gamma\gamma}$ coupling~\cite{Su:2019ibd} once $\kappa_t$ flips sign.  Therefore, no such wrong-sign region appears in the lower-left corner of the plot.   Note that there is also an allowed region for Type-I appearing in the upper left corner at one-loop level, which corresponds to $\kappa_{f/W/Z}=-1$ caused by loop corrections.

Comparing the one-loop results (solid curves) with that of the tree-level (dashed curves) in \autoref{fig:lhcrun2limit}, we see that other than the large $\tan\beta$ region of Type-I~\cite{Chen:2019pkq} mainly due to loop corrections to $hZZ$ coupling, the impact of loop corrections on the allowed region in $\tanb$-$\cosba$ plane is small.  The ``wrong-sign" region is shifted at one-loop for relatively large $\lambda v^2$ since the tree level shift is close to zero in this region.

\begin{figure}[!tbp]
\centering
\includegraphics[width=0.45\textwidth]{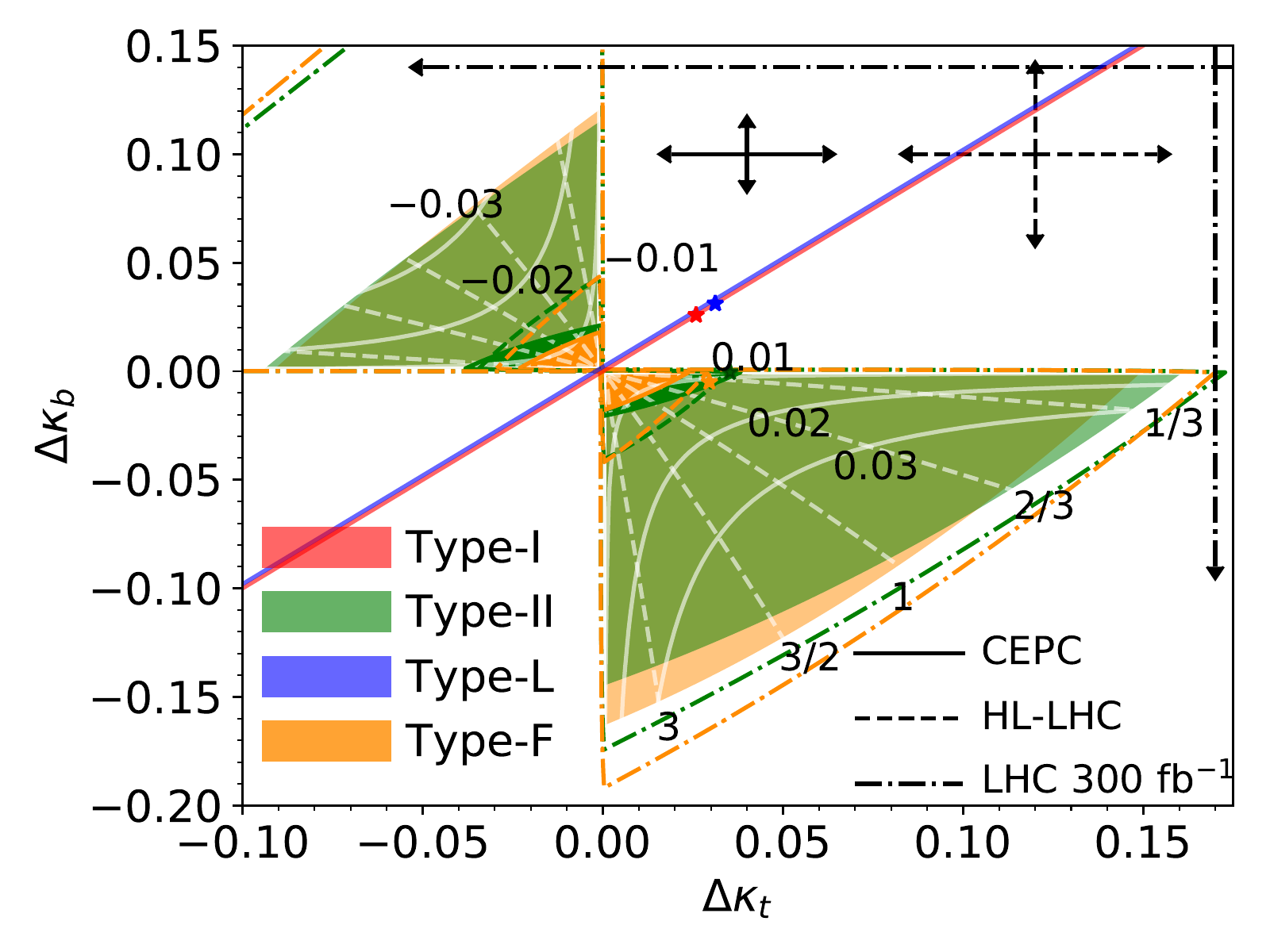}
\includegraphics[width=0.45\textwidth]{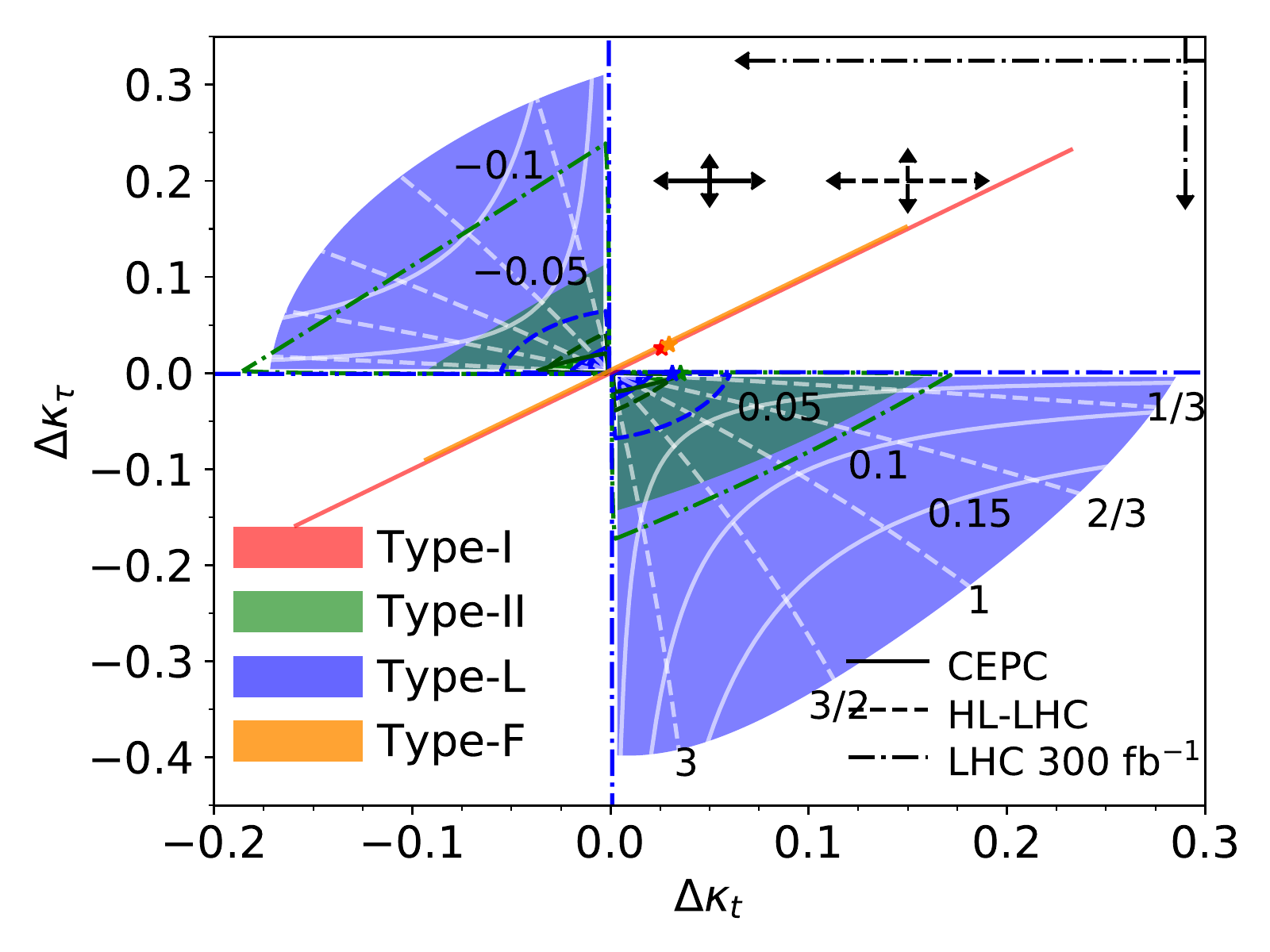}
\includegraphics[width=0.45\textwidth]{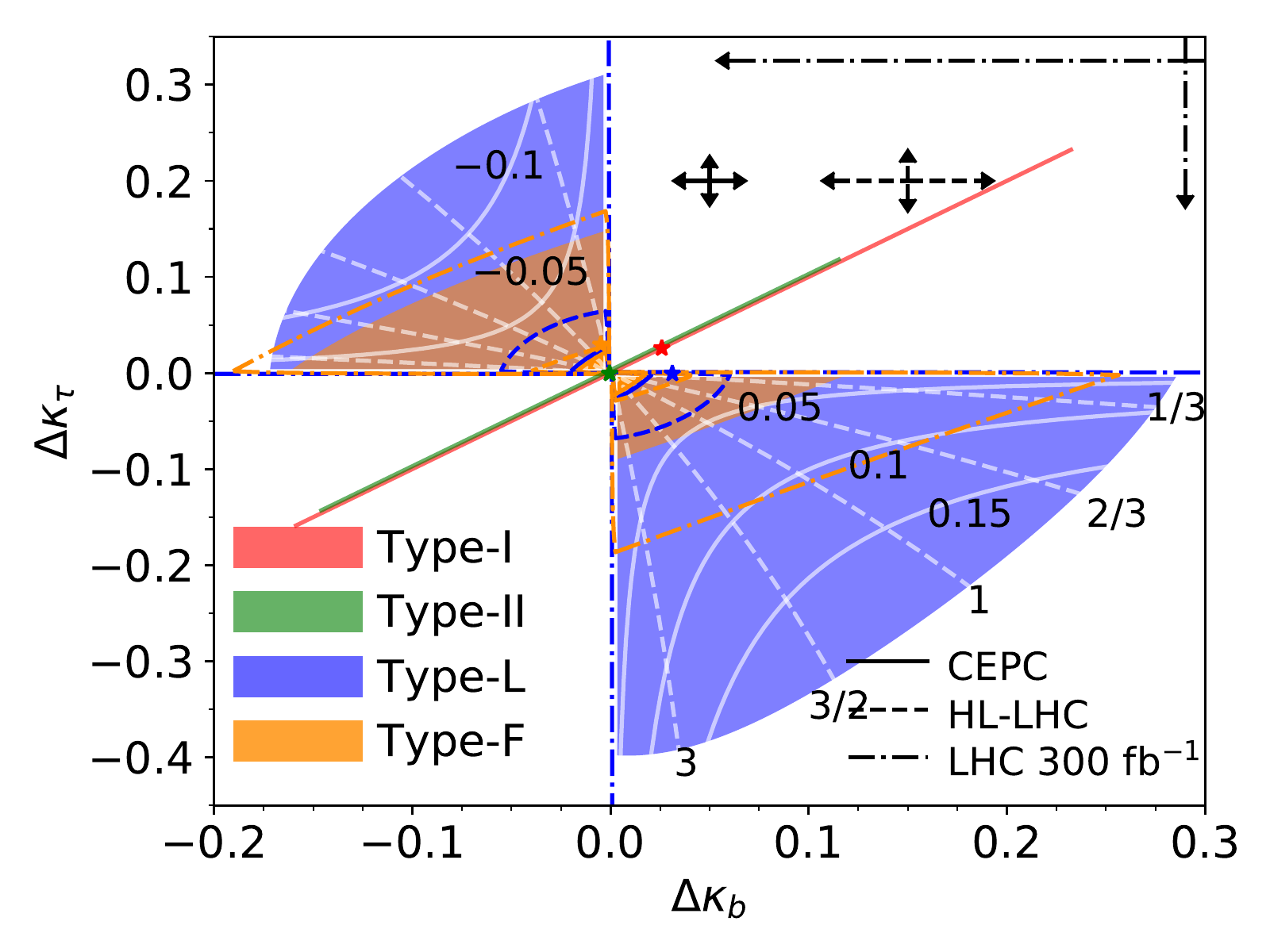}
\includegraphics[width=0.45\textwidth]{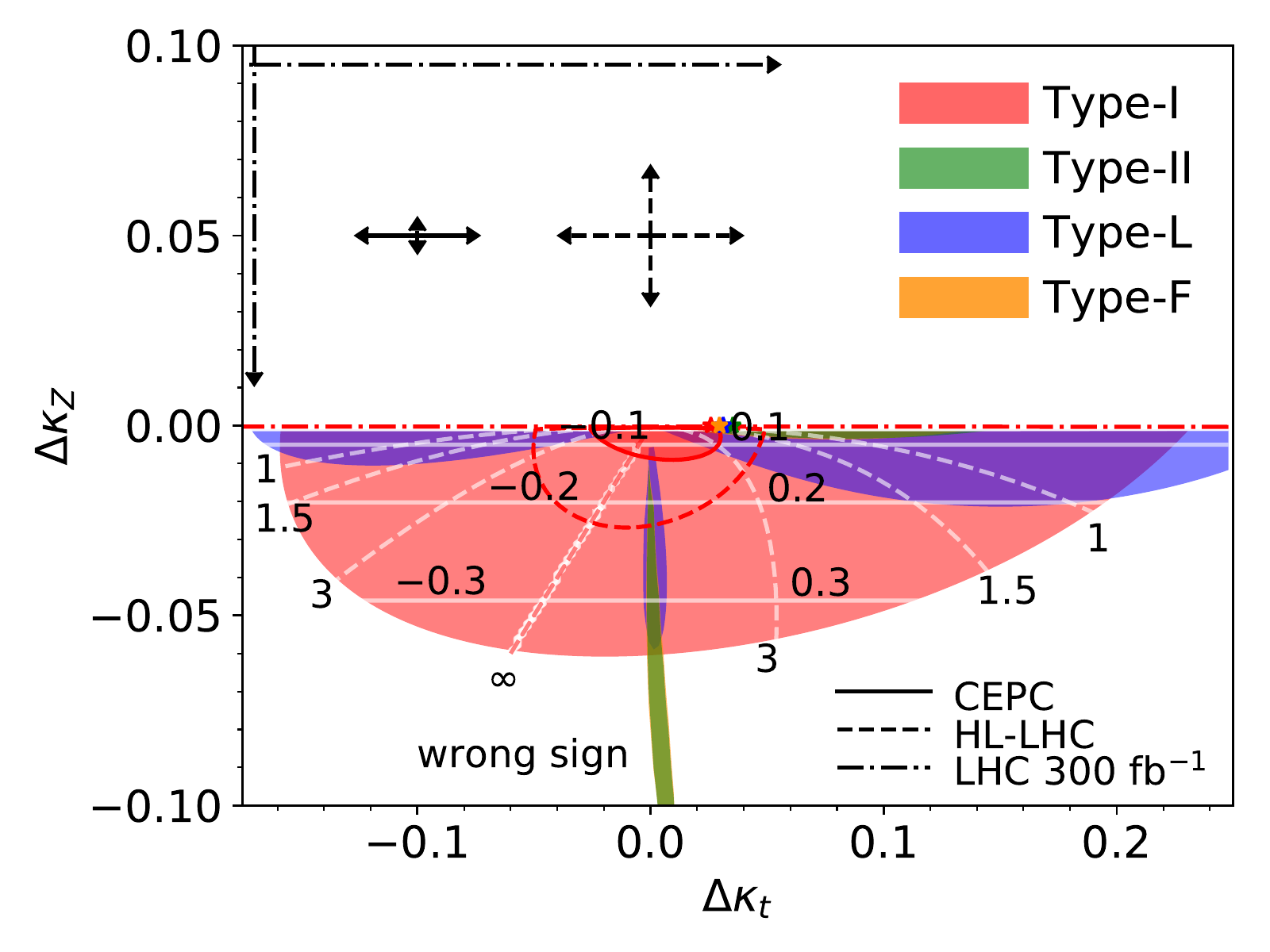}
\includegraphics[width=0.45\textwidth]{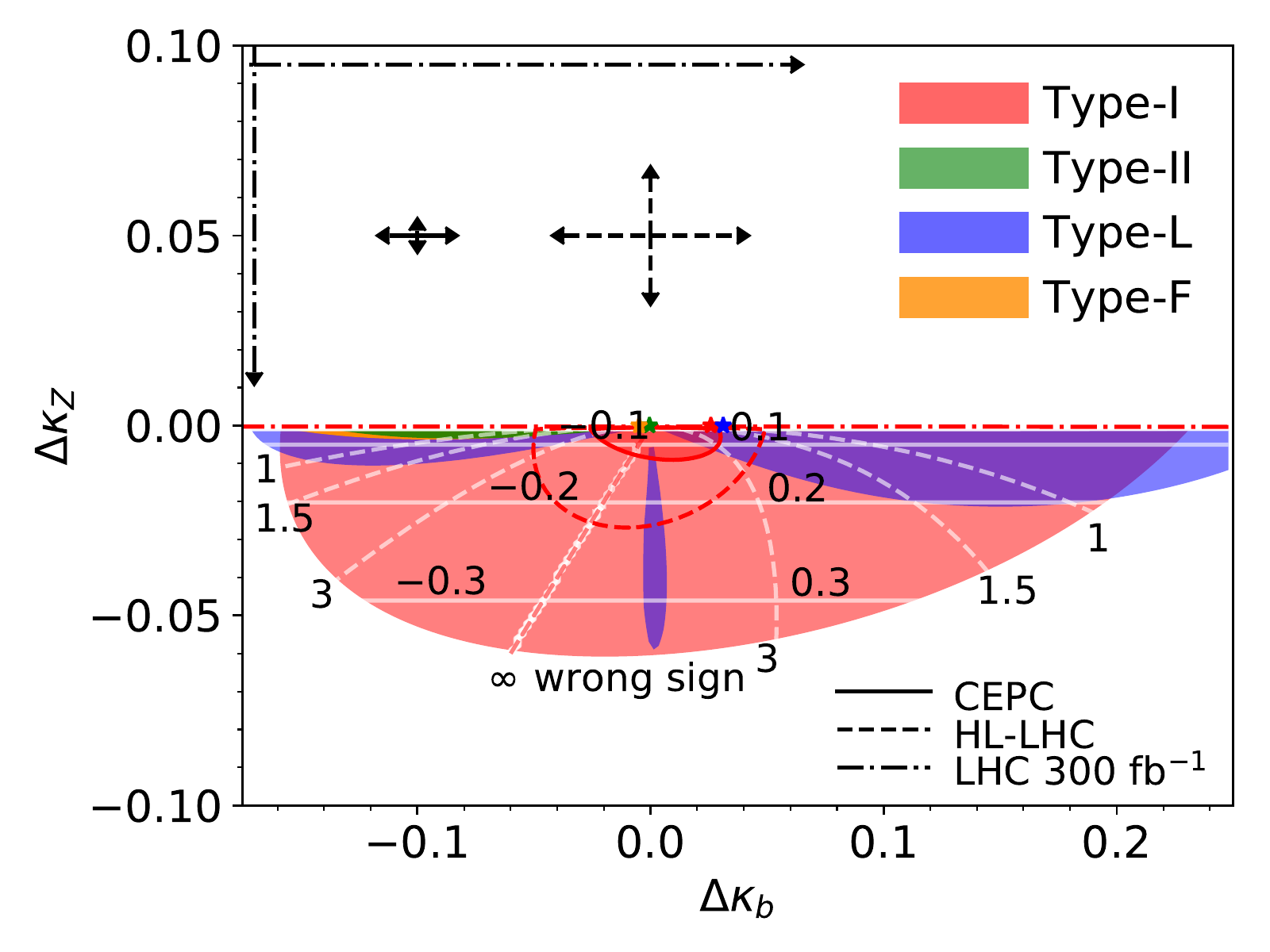}
\includegraphics[width=0.45\textwidth]{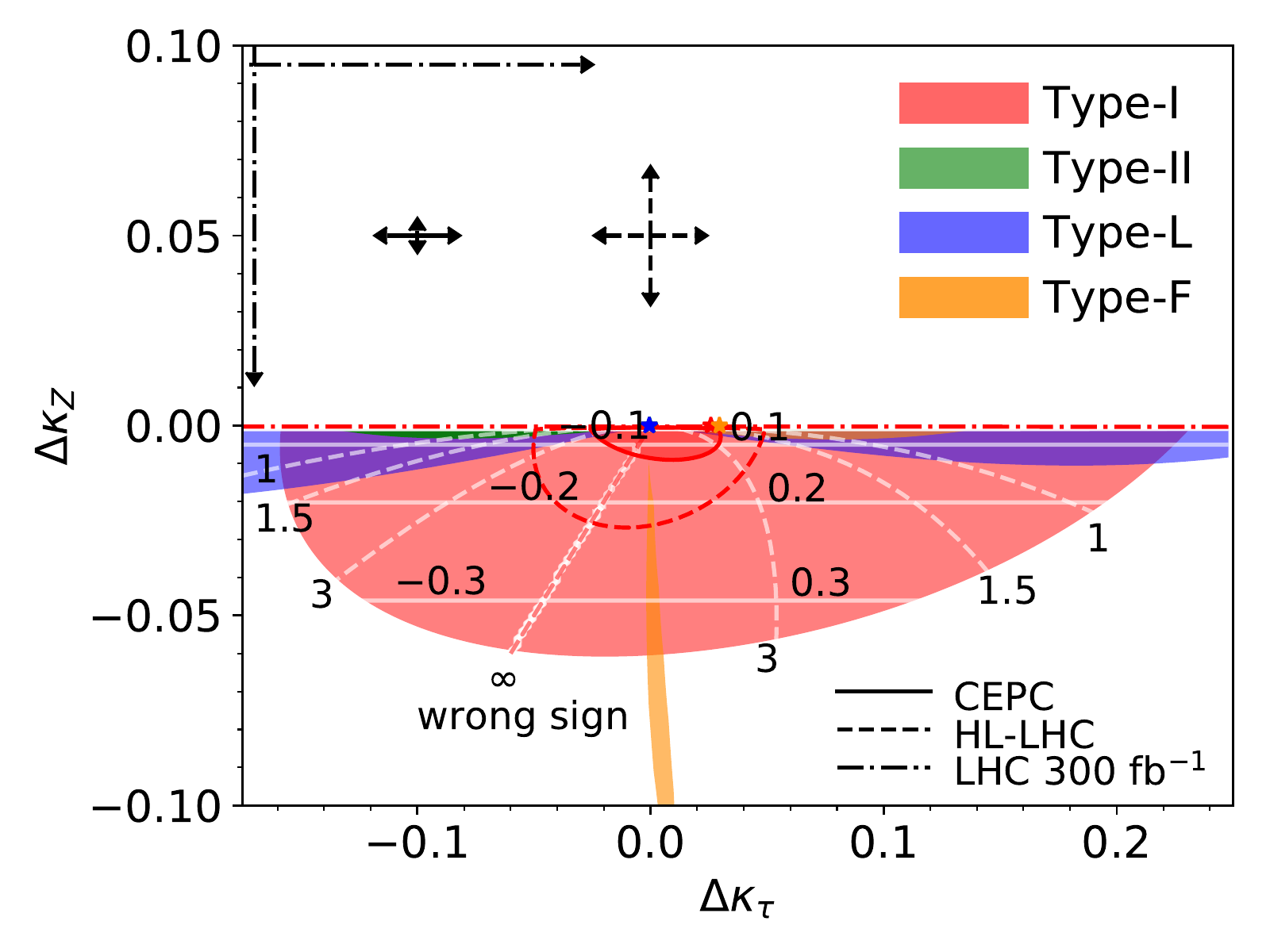}
\caption{95\% C.L. allowed regions in the  $\Delta \kappa_i$-$\Delta \kappa_j$ plane  ($i,j=t(c), b, \tau, Z$) from the tree-level results under the current LHC limits. Results of Type-I, II, L and F are indicated by the red, green, blue and orange colors, respectively. The regions outside the solid, dashed and dash-dotted lines in the corresponding colors indicate the $5\sigma$ discovery reaches of CEPC, HL-LHC and LHC (300$\fb^{-1}$).  }
\label{fig:lhcrun2limitkappa}
\end{figure}

To compare with experimental observations,  we map out the allowed region in 2HDM parameter space under the current LHC limit to the deviations in various couplings normalized to the SM value: $\Delta\kappa_i\equiv\kappa_i-1$, for $i=t,b,\tau$ or $Z$.  Given that different types of 2HDMs predict different Yukawa coupling relations, we present the results in the  $\Delta\kappa_i$-$\Delta\kappa_j$ plane. 
 
To better understand the qualitative features, we present the tree-level  results first in \autoref{fig:lhcrun2limitkappa}   in the  $\Delta \kappa_i$-$\Delta \kappa_j$ plane.  The shaded regions display the LHC allowed regions for four types of 2HDMs, with red, green, blue and orange colors referring to Type-I, II, L and F, respectively.   The LHC best fitting points for $\Delta\kappa$ are marked by  stars in the corresponding colors.  The solid, dashed and dash-dotted lines exhibit the discovery reach of the CEPC, HL-LHC and LHC of 300$\fb^{-1}$ luminosity, outside which a discovery of $5\sigma$ significance or above could be made. The projected experimental precision on measuring $\kappa_i$ at different machines is indicated by crossing arrows with the consistent line styles.  Also shown are the values of $\cos(\beta-\alpha)$ and $\tan\beta$ by solid and dashed white contour lines. For the two upper panels, the overlapping two types in the second and fourth quadrants share the same white contour lines, whereas for the middle left panel, Type-L and Type-F have the opposite sign in $\cos(\beta-\alpha)$: the labeled values are for Type-L. The white contours in the last three panels are for Type-I specifically. For the  CEPC curves, $\Delta\kappa_c$ is used instead of $\Delta\kappa_t$ given the same Yukawa coupling structure and better experimental precision expected for the charm Yukawa coupling at the CEPC (see \autoref{tb:exp_kappa}).  

In the first three panels in \autoref{fig:lhcrun2limitkappa}, the Type-I results show up as straight  diagonal lines, since   $\kappa_t=\kappa_b=\kappa_\tau$ at tree level.
Similarly, $\kappa_t=\kappa_b$ in Type-L, $\kappa_t=\kappa_\tau$ in Type-F, and $\kappa_b=\kappa_\tau$ in Type-II, which are reflected in the diagonal lines in the first three panels as well.   While no correlation appears for $\Delta\kappa_t$ and $\Delta\kappa_b$ in Type-II and Type-F, they only occupy the second and fourth quadrants as shown in the upper-left panel.  The allowed region appears asymmetric because the central values of the current LHC data deviate slightly away from the SM prediction.  Similar behaviour for $\Delta\kappa_t$ vs. $\Delta\kappa_\tau$ and $\Delta\kappa_b$ vs. $\Delta\kappa_\tau$ can be observed in the upper-right and middle-left panel as well for different types of 2HDMs.

The remaining three panels show  $\Delta\kappa_Z$ versus  $\Delta\kappa_{t,b,\tau}$, with only Type-I extending much further toward negative $\Delta\kappa_Z=\sin(\beta-\alpha)-1$ direction than the other types. This is because 
Type-I permits much wider range of $\cos(\beta-\alpha)$ away from the alignment limit, as shown in  \autoref{fig:lhcrun2limit}. The best fitting points in the last three panels stick to the upper boundary of the allowed regions. This is because the current LHC experiments yield  positive $\Delta\kappa_Z=0.1 \pm 0.08$~\cite{Aad:2019mbh}, whereas $\Delta\kappa_Z$ in 2HDMs is always negative. 
The narrow bands around $\Delta\kappa_{t(c),b,\tau}=0$ of Type-II, L and F 
are the wrong sign regions, which confirm that the corrections to the $\Phi_2$ induced Yukawa couplings vanish in this very region.

Comparing  the LHC allowed region with the $5\sigma$ discovery reach of the future measurements, we can see that almost the entire allowed region permits a discovery at the CEPC and HL-LHC, whereas the discovery reach of LHC of 300 $\fb^{-1}$ integrated luminosity is rather limited. Combining all six panels, one finds that the four types of 2HDMs exhibit very distinct distributions in $\kappa$-space except for the common intersection at origin. For instance, Type-I and L overlap in $\Delta\kappa_t$-$\Delta\kappa_b$ plane, while they are  completely separable from Type-II and F. To further distinguish Type-I and L, one can examine $\Delta\kappa_\tau$  in addition, as shown in the top-right and middle-left panels.  Measuring $\Delta\kappa_Z$ also provides an immediate separation between Type-I and the other three types, if a large deviation is observed. Therefore, if a deviation of the Higgs couplings from the SM prediction is observed at future experiments,  one could potentially  distinguish the four different types of 2HDMs by measuring all the four couplings.  

\begin{figure}[!tbp]
\centering
\includegraphics[width=0.45\textwidth]{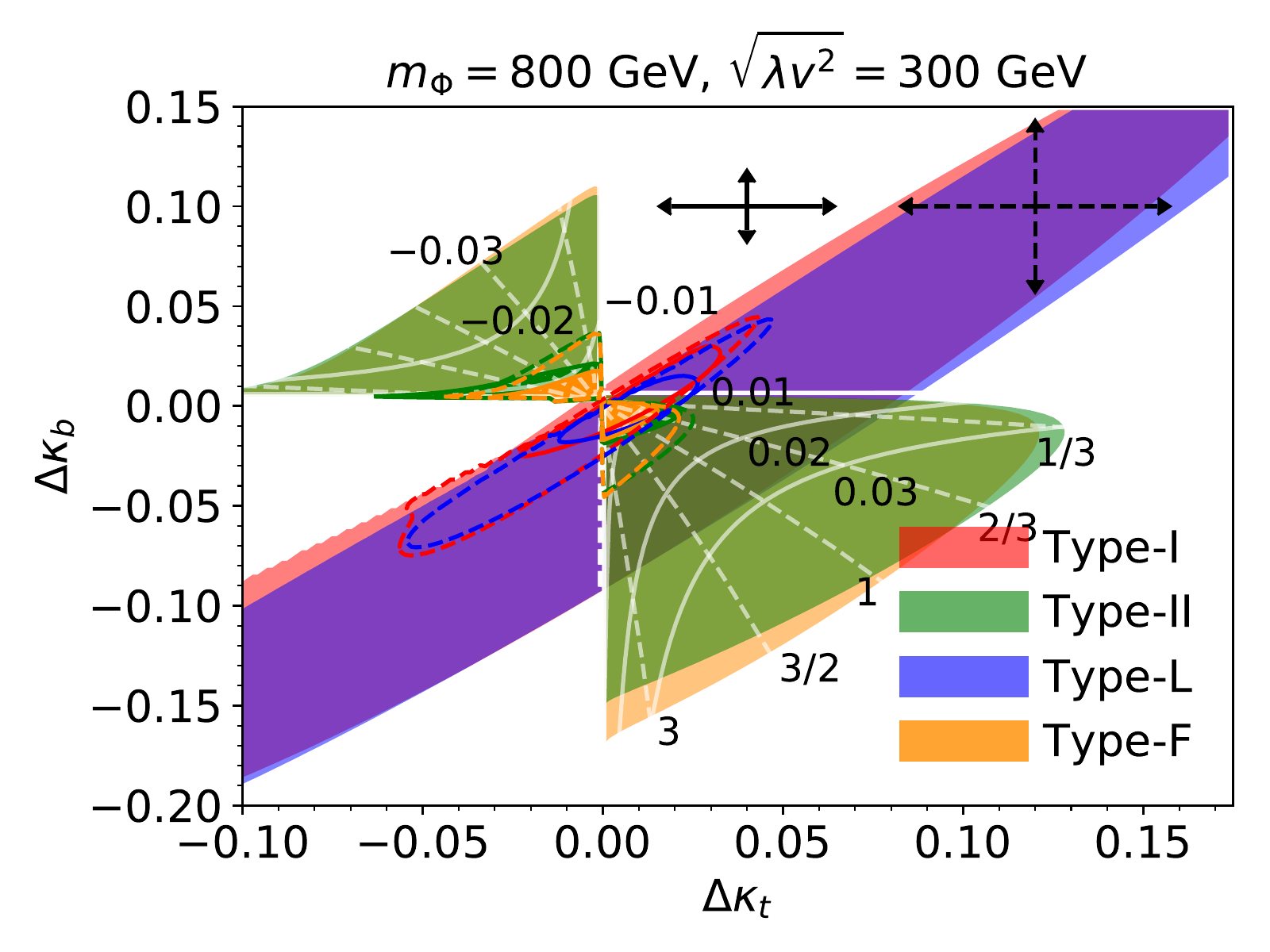}
\includegraphics[width=0.45\textwidth]{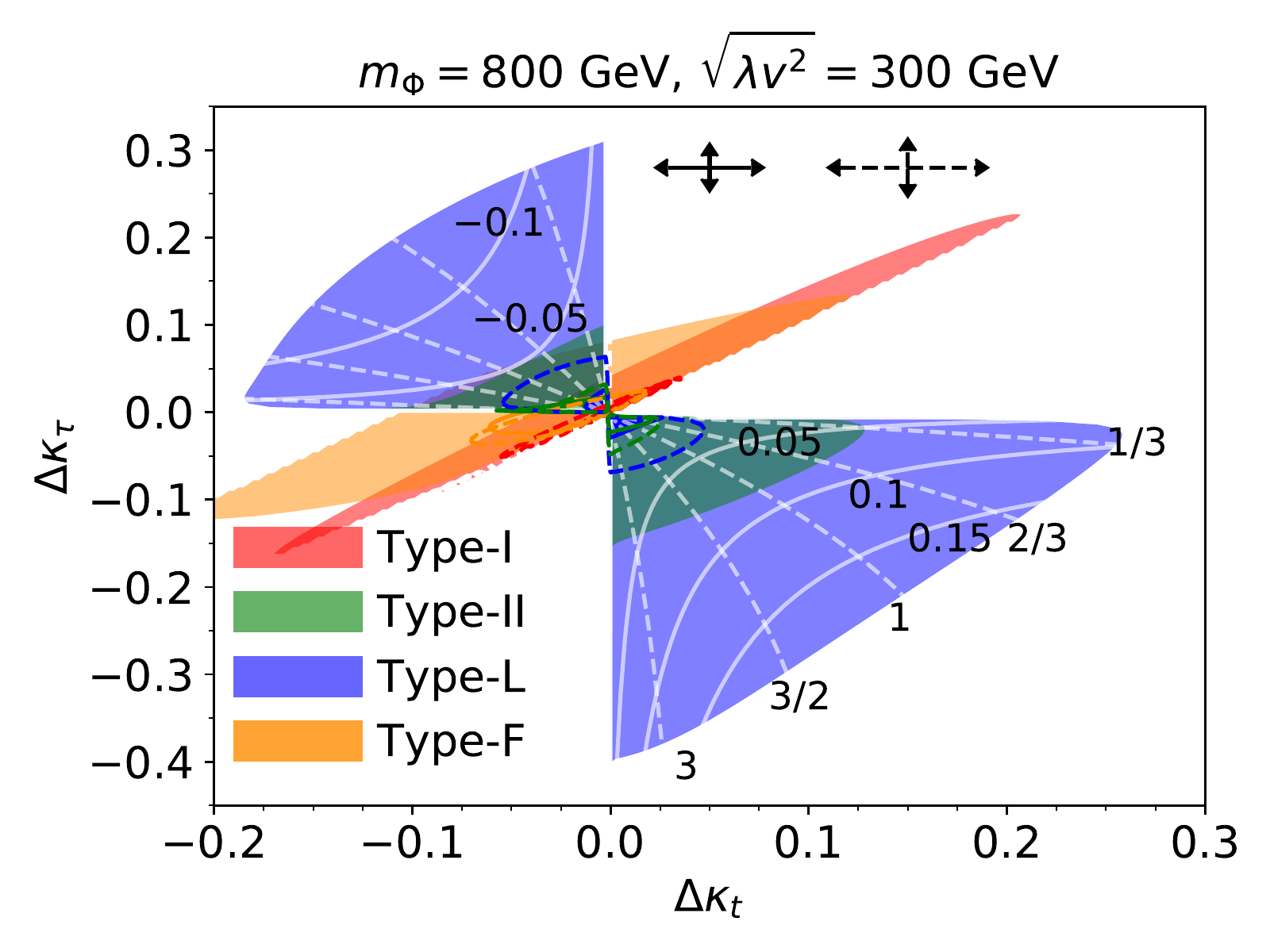}
\includegraphics[width=0.45\textwidth]{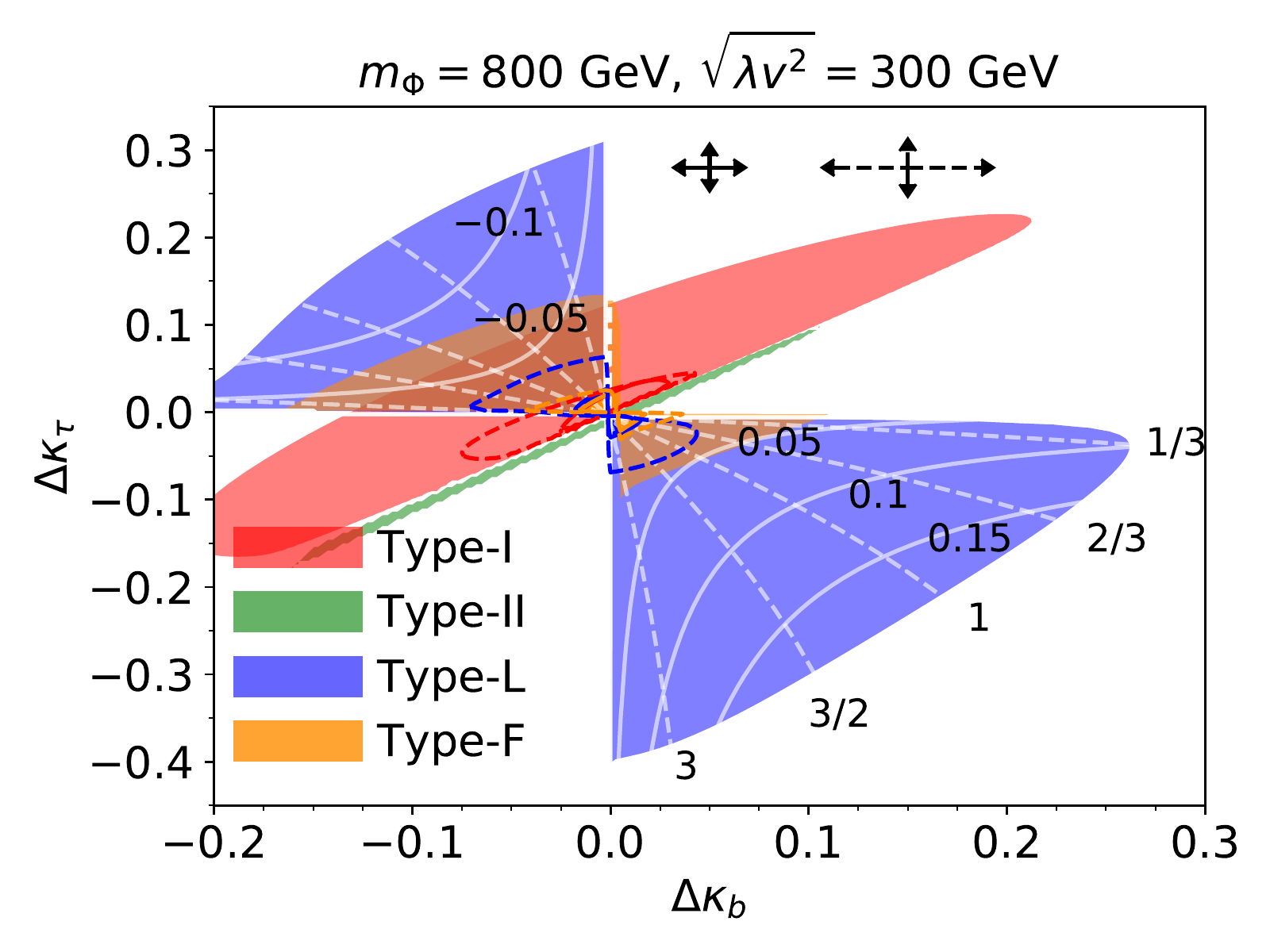}
\includegraphics[width=0.45\textwidth]{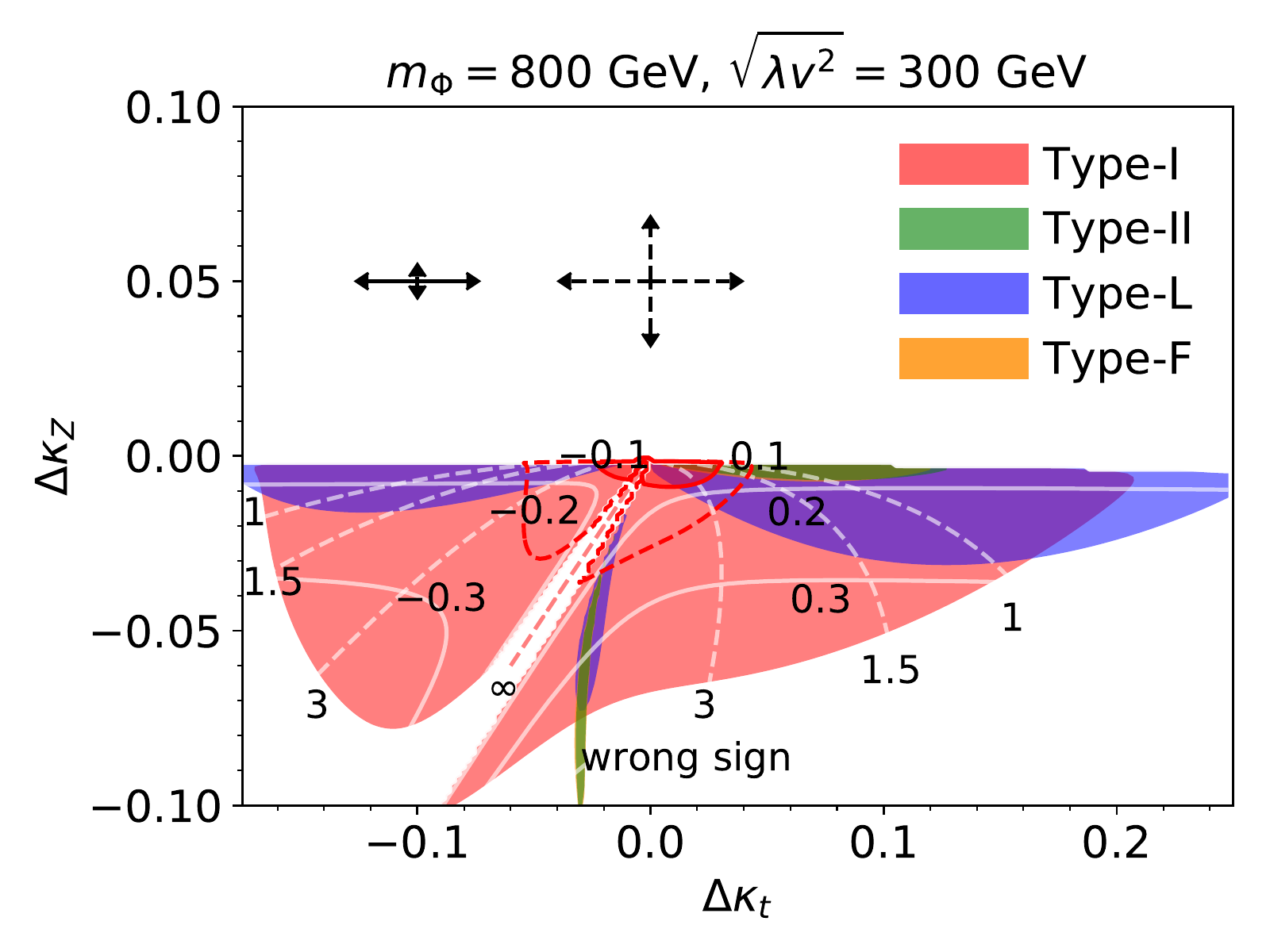}
\includegraphics[width=0.45\textwidth]{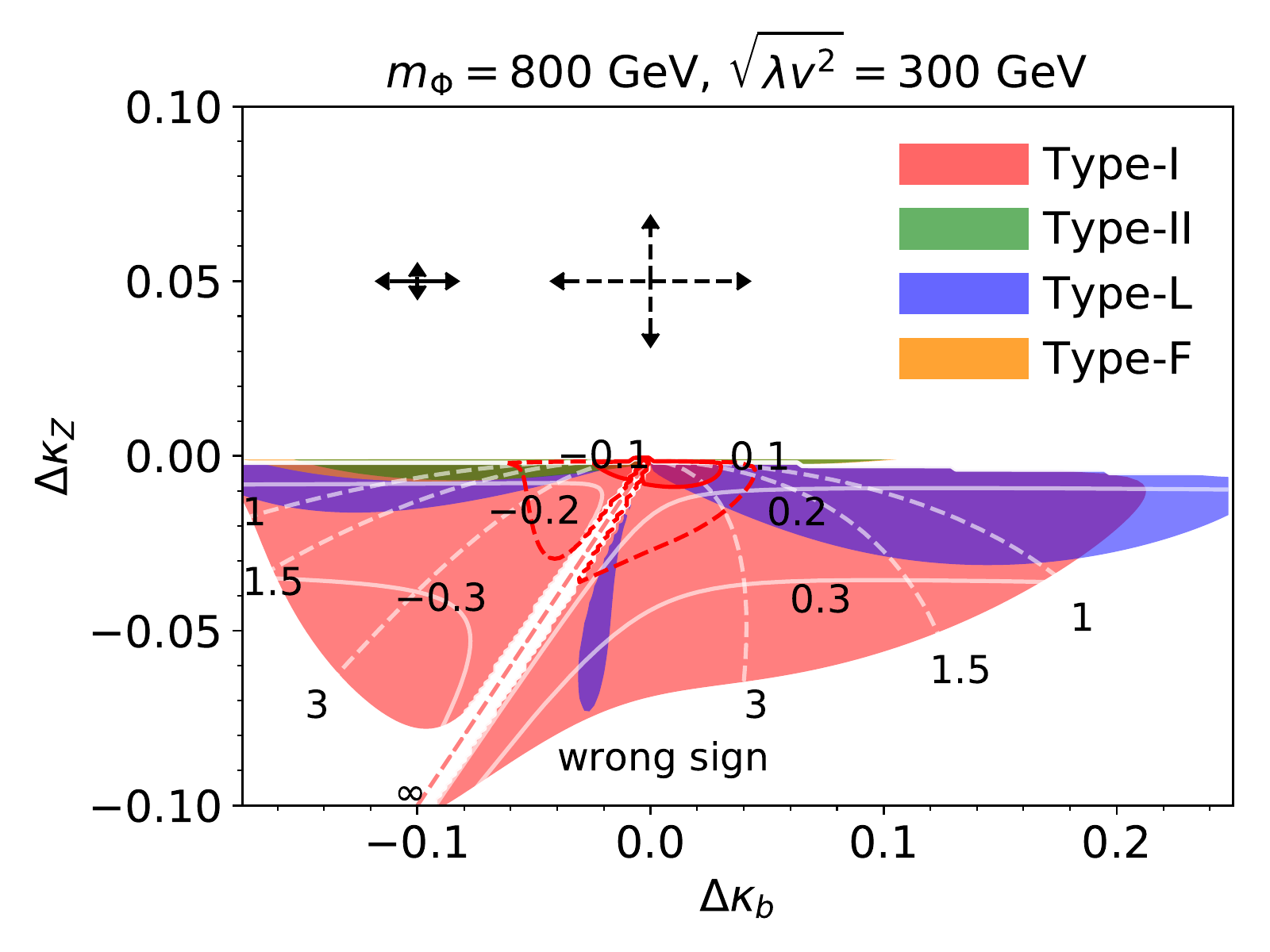}
\includegraphics[width=0.45\textwidth]{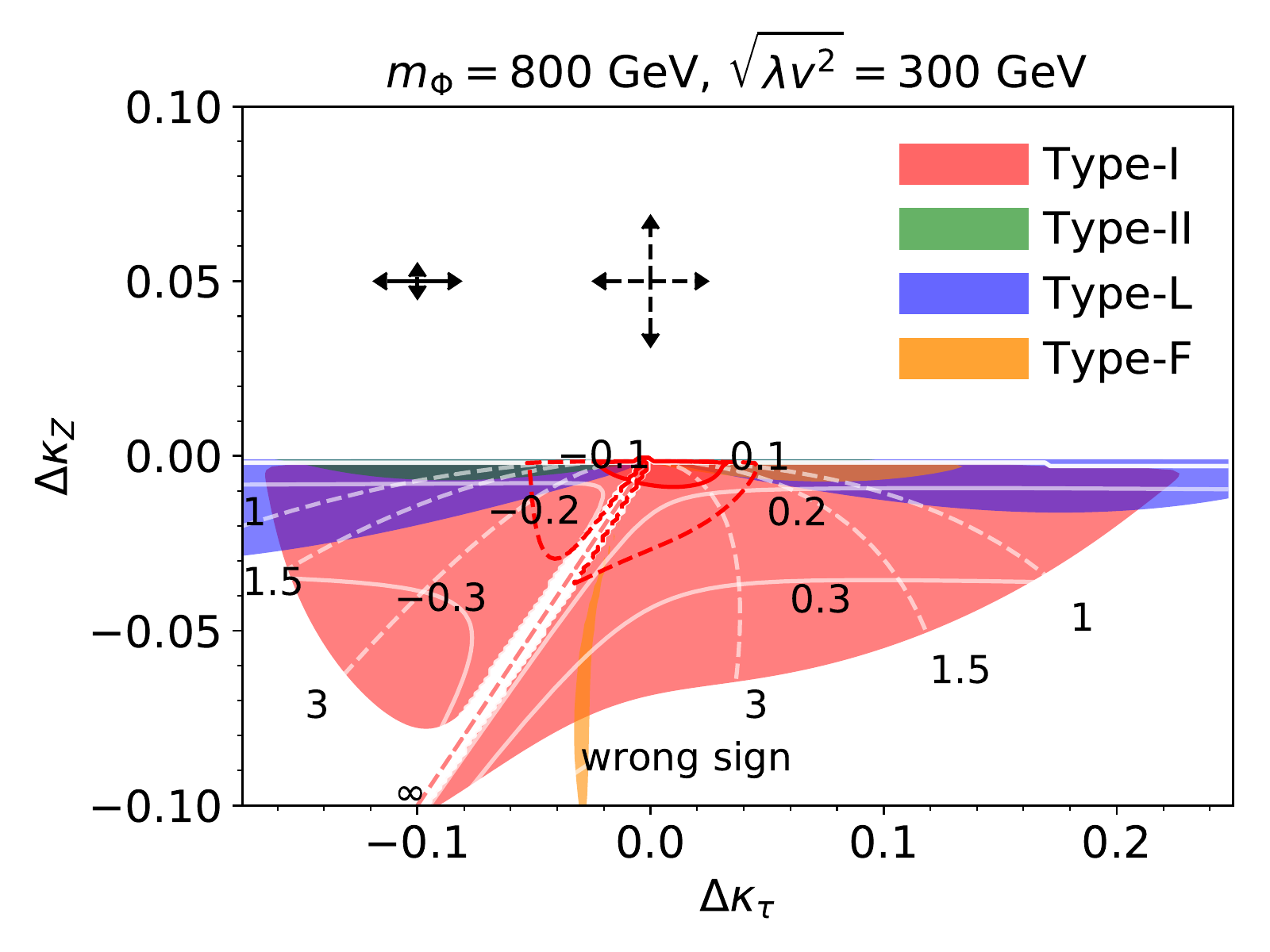}
\caption{95\% C.L. allowed regions in the  $\Delta \kappa_i$-$\Delta \kappa_j$ plane  ($i,j=t(c), b, \tau, Z$), the same as \autoref{fig:lhcrun2limitkappa}, but from the one-loop results.}
\label{fig: lhcrun2limitkappaloop_300}
\end{figure}

However, once loop corrections are included, the correlations among couplings are smeared.  For illustration, we show in \autoref{fig: lhcrun2limitkappaloop_300} the LHC allowed regions as well as the future discovery reaches at one-loop with degenerate non-SM Higgs mass $m_\Phi=m_H=m_A=m_{H^\pm}=800$ GeV and $\sqrt{\lambda v^2}=300$ GeV.  The tree-level relation of $\kappa_{t(c)}$, $\kappa_b$ and $\kappa_\tau$  receives loop corrections, which result in the wide bands along the diagonal direction. The linear correlation between $\kappa_b$ and $\kappa_\tau$ for Type-II still persists (middle-left panel), due to the relatively small loop corrections for  both the bottom and tau Yukawa couplings in the Type-II 2HDM.

In the last three panels of  \autoref{fig: lhcrun2limitkappaloop_300},  the allowed region of Type-I is seen to be distorted by loop corrections significantly  and the wrong-sign regions are shifted to the left compared to the tree-level results. Since the wrong-sign regions always stay in the positive $\cos(\beta-\alpha)$ region (see \autoref{fig:lhcrun2limit}), they are truncated at the $\tan\beta \to \infty$ contour line.

While the loop corrections weaken the capability to distinguish different types of 2HDMs,  in particular due to the spread of the diagonal regions in Yukawa coupling correlations, combining all six panels still demonstrates the advantage in discrimination.  A quantitative method to distinguish four different types of 2HDMs is given in the next section, by utilizing the $\chi^2$ distribution.  

We present the CEPC $5\sigma$ discovery contour lines in \autoref{fig:loopvstree5sigma} in $\cos(\beta-\alpha)$-$\tan\beta$ plane, at tree (dashed) and one-loop (solid) level. Red, green, blue and orange colors correspond to Type-I, II, L and F, respectively.  The regions outside the contour lines are accessible with $5\sigma$ sensitivity.  A degenerate scalar mass of $m_H=m_A=m_{H^\pm}=m_\Phi=800$ GeV and $\sqrt{\lambda v^2}=300$ GeV is used for obtaining the one-loop results.   For the Type-I 2HDM, regions with $\cos(\beta-\alpha)\lesssim -0.1$ or $\cos(\beta-\alpha)\gtrsim 0.08$ are discoverable at more than $5\sigma$ level.  For the other three types of 2HDMs, the $5\sigma$ region is even bigger: $|\cos(\beta-\alpha)|\gtrsim 0.02$ for $\tan\beta\sim 1$.   At small and large values of $\tan\beta$, the region in $\cos(\beta-\alpha)$ is further tightened. 

\section{Distinguishing the Four Types of 2HDMs}
\label{sec:impl}

\begin{figure}[!tbp]
\centering
\includegraphics[width=0.6\textwidth]{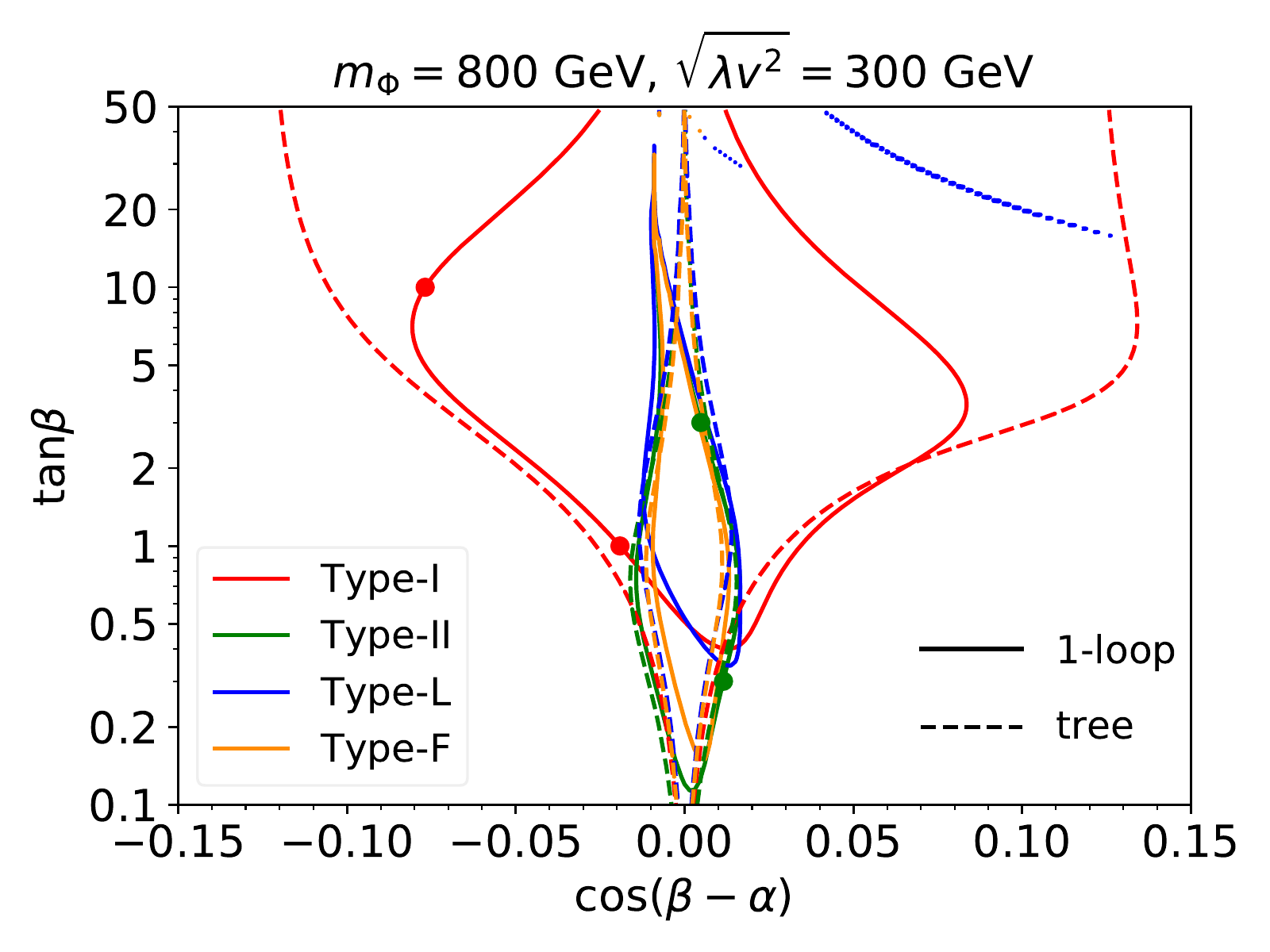}
\caption{$5\sigma$ discovery regions outside the contour lines in the $\cos(\beta-\alpha)$-$\tan\beta$ plane  for CEPC at tree level (dashed) and one-loop (solid). Red, green, blue and orange colors indicate the results of Type-I, II, L and F, respectively. We choose $m_H=m_A=m_{H^\pm}=m_\Phi=800$ GeV and $\sqrt{\lambda v^2}=300$ GeV for one-loop curves. Four representative points in Type-I and II are marked with red and green dots, respectively.} 
\label{fig:loopvstree5sigma}
\end{figure}

It is encouraging to see the discovery potential at the $5\sigma$ level from the precision SM measurements and to realize the characteristic features of different types of 2HDMs as shown in the last section. We now make a few case studies to quantify the feasibility to distinguish the four different types of 2HDMs. Our procedure is that we start with a benchmark point in the 
$\cos(\beta-\alpha)$-$\tan\beta$ plane that permits a $5\sigma$ discovery at one-loop level with the precision of CEPC for a particular type of 2HDM, called the target model. We study how the target model can be distinguished from other types  with a quantitative $\chi^2$ analysis. In particular, we take the corresponding $\mu_i$ for that benchmark point as $\mu_i^1$ in \autoref{eq:chi2}, and perform a $\chi^2$ analysis with $\mu_i^0$ being the signal strength of the other 2HDMs. A $2\sigma$ significance of inconsistency  is set as the criterion for the model  discrimination, which corresponds to roughly 95\% C.L.

We choose four benchmark points for two target models of Type-I (red dots) and Type-II (green dots) in \autoref{fig:loopvstree5sigma} at a  small and large value of $\tan\beta$ to perform the comparative study.  The values of $\cos(\beta-\alpha)$ and $\tan\beta$ for the benchmark points labelled as IA, IB and IIA, IIB, are summarized in \autoref{tb:coord}.

\begin{table}[!tbp]
\centering
\begin{tabular}{c|c c}
\hline 
\hline
 $(\cos(\beta-\alpha),\tanb)$&Small $\tanb$ &Large $\tanb$ \\
\hline
Type-I & {\bf{IA:}} ($-$0.019,1.0) &{\bf{IB:}} ($-$0.077,10)\\
\hline
Type-II & {\bf{IIA:}} (0.012,0.3) & {\bf{IIB:}} (0.005,3.0)\\
\hline
\hline
\end{tabular} 
\caption{Benchmark points of $\cos(\beta-\alpha)$ and $\tan\beta$ in Type-I and Type-II 2HDMs with a low and high value of  $\tanb$.}
\label{tb:coord}
\end{table}

\begin{figure}[!tbp]
\centering
\includegraphics[width=0.49\textwidth]{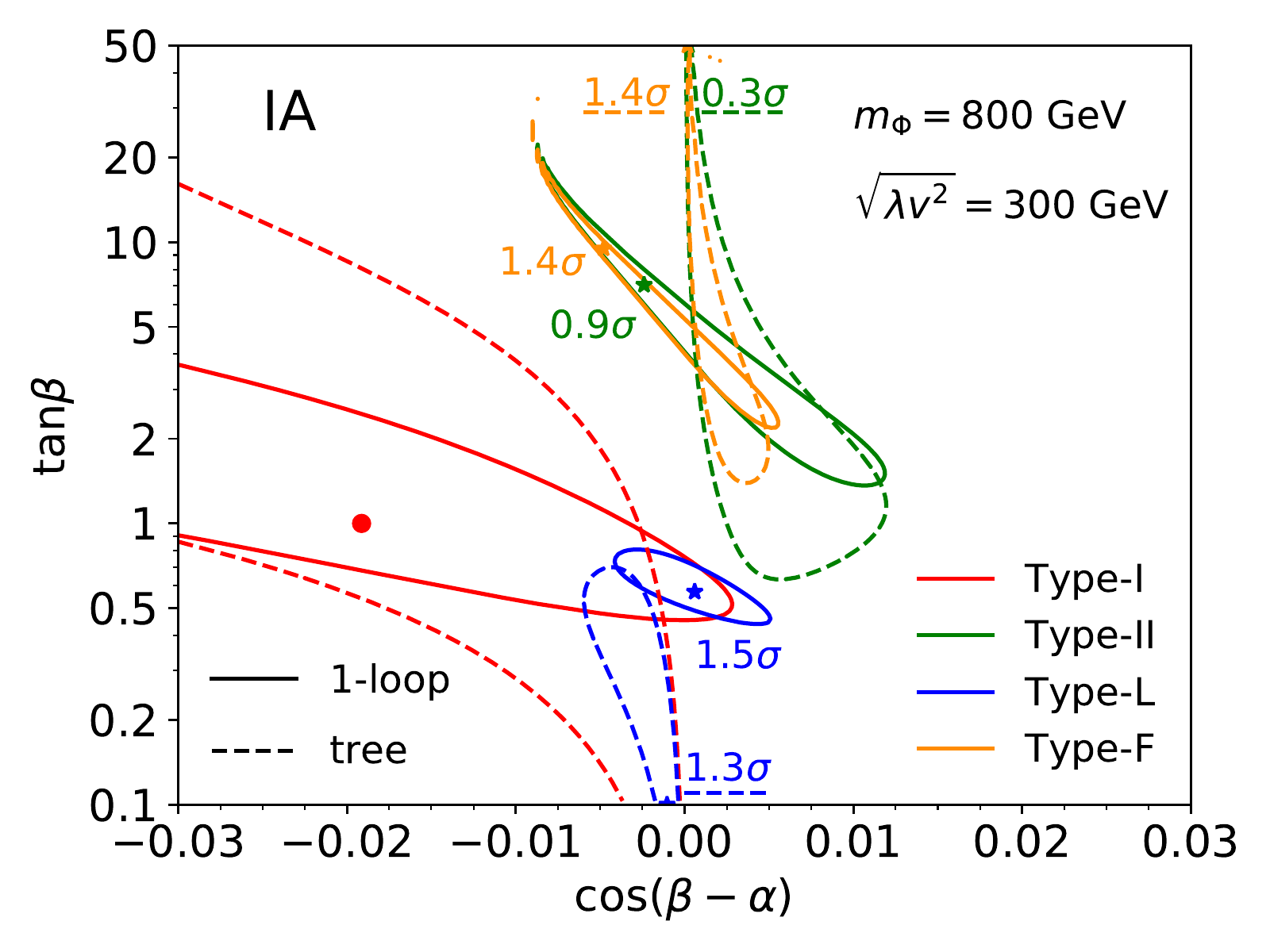}
\includegraphics[width=0.49\textwidth]{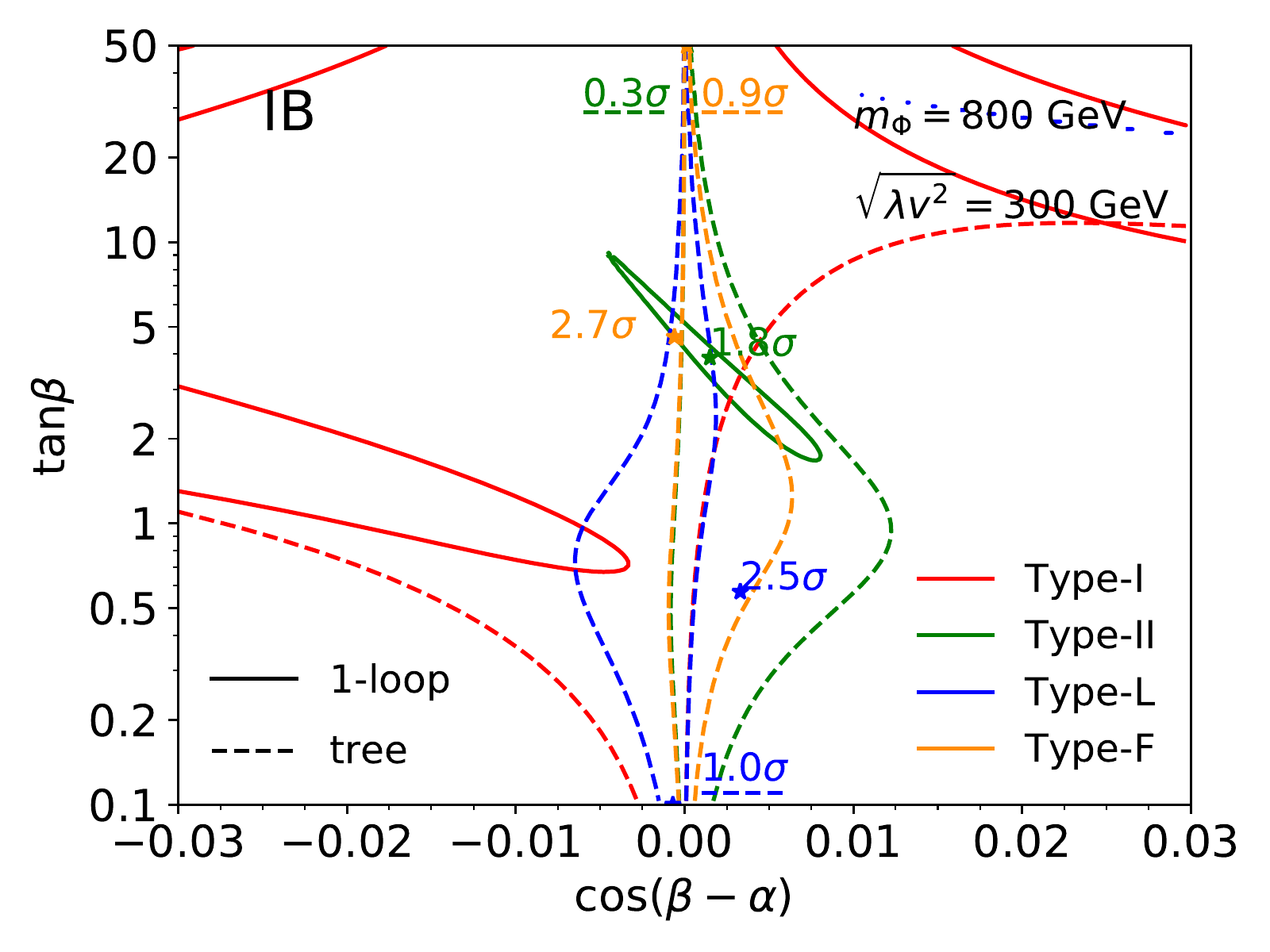}
\includegraphics[width=0.49\textwidth]{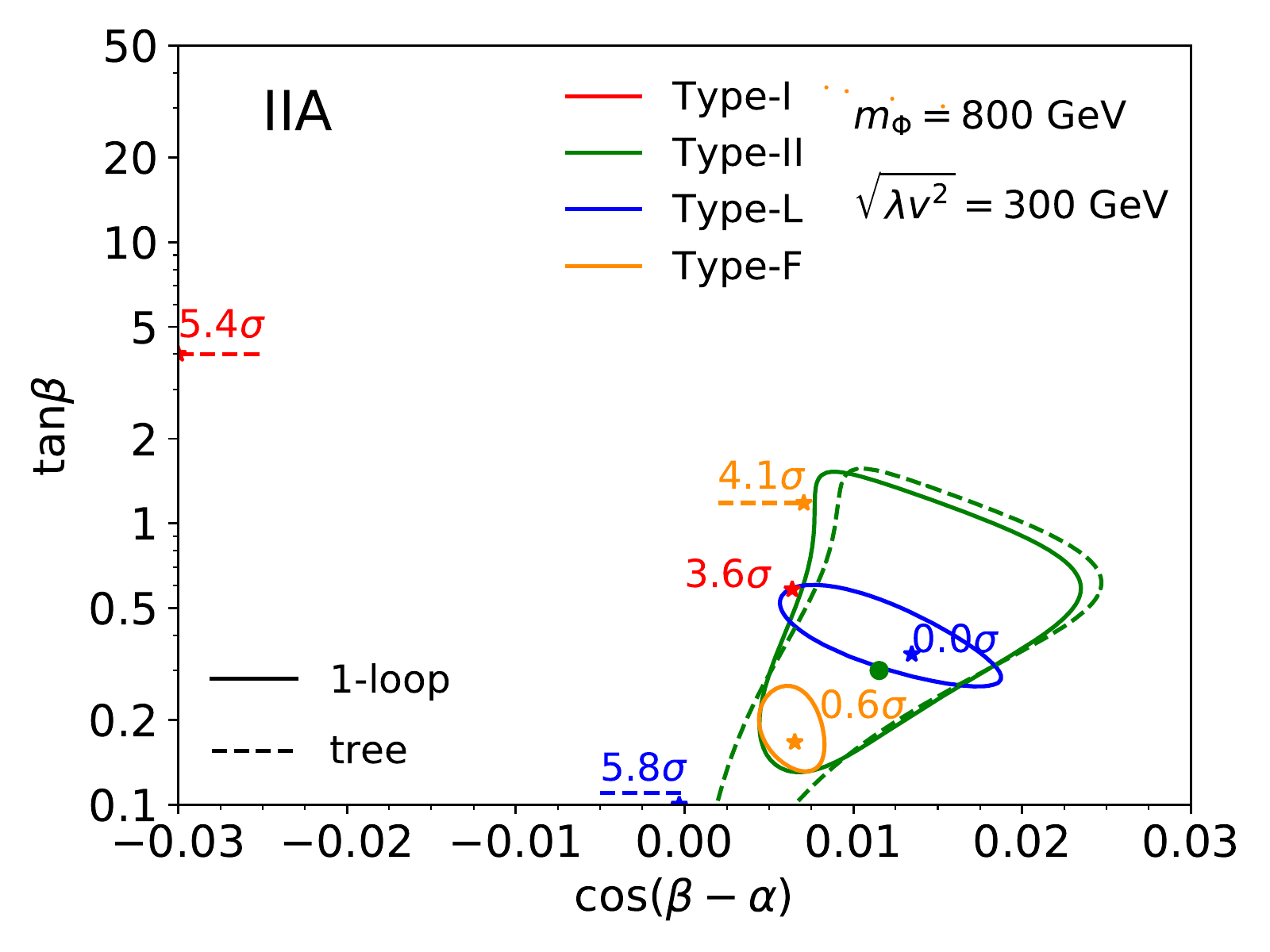}
\includegraphics[width=0.49\textwidth]{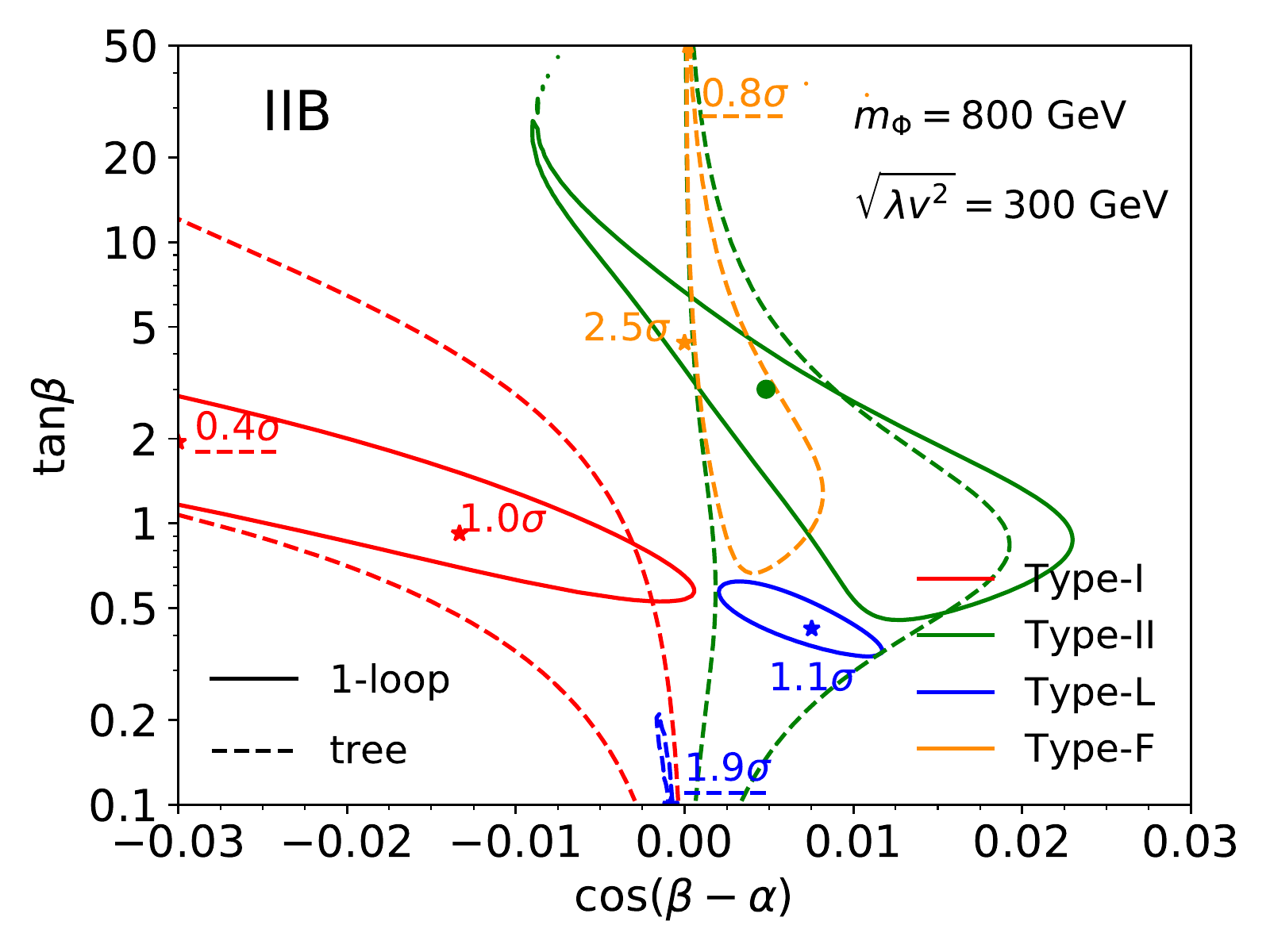}
\caption{95\% C.L. discrimination regions with solid (dashed) contours from one-loop (tree level) results with respect to the benchmark points (dots) of Type-I (upper panels) and Type-II (lower panels).  Red, green, blue and orange colors refer to the points in Type-I, II, L and F, respectively. 
Stars of consistent colors mark the best fitting point, of which the significance of inconsistency are labelled beside.      }
\label{fig:compareloop}
\end{figure}

In \autoref{fig:compareloop}, we show the 95\% C.L.  discrimination regions of different types of 2HDMs from the benchmark points of the target models in the $\cos(\beta-\alpha)$-$\tan\beta$ plane. The top (bottom) two panels are for the benchmark points in Type-I (Type-II), and  the left and right panels are for small and large $\tan\beta$ benchmark points, respectively.  In each panel, the benchmark point is marked with small dots of the respective color, unless it lies outside the plot range as in the upper-right panel. The solid contour lines indicate the 95\% C.L.  distinguishable regions outside the contours at one-loop, with the best fitting points indicated by the stars of the corresponding colors. Also shown is corresponding significance for the best fitting point. Regions within the contours are the parameter space of the corresponding 2HDM type that cannot be distinguished from the target model of the benchmark point at 95\% C.L.  Note that lines of identical color to the benchmark point enclose the region that is consistent with the benchmark point of the target model at 95\% C.L.   In this process, we do one test for each point comparing with the benchmark point, thus the number of d.o.f equals the number of SSMs, which is ten in CEPC.

For the small $\tanb$ Benchmark IA of Type-I shown in the upper-left panel, the best fitting point of Type-II 2HDM is consistent at  $0.9\sigma$ while the region enclosed by the green contour of Type-II cannot be distinguished from Benchmark IA at 95\% C.L. Similar conclusion holds for Type-L (blue) and Type-F (orange). To see the impact of loop corrections, we also present the $\chi^2$-fit results at tree level, which are shown via dashed contour lines, with numbers underlined with dashed lines being the best fitting point significance at tree level.  Including loop corrections shifts the 95\% C.L. region for all the three types considerably. 

For the large $\tanb$ Benchmark IB in Type-I shown in the upper-right panel, other than a small slice of the region for the Type-II 2HDM, all the other three types can be distinguishable from Benchmark IB at 95\% C.L.  The loop corrections are more significant here, as the dashed regions at tree level disappear once loop effects are included. 
Note that the red curve indicates the region of the Type-I 2HDM that is consistent with the Benchmark IB, which contains two regions between two red solid curves in the upper left and upper right corners, as well as the region enclosed by the red solid curve around $\tan\beta\sim 2$.

 For the small $\tanb$ Benchmark IIA in Type-II shown in the lower-left  panel, while it can be easily separable from Type-I, the best fitting point in Type-L is very close to Benchmark IIA with $\chi^2=0$. As a direct consequence of loop corrections, this is caused by the strong overlap between Type-II and L in $\kappa$ space displayed in the first panel of \autoref{fig: lhcrun2limitkappaloop_300}.  No tree-level dashed region appears for this benchmark point, again showing the effect of loop corrections.

 For the large $\tanb$ Benchmark IIB point in Type-II shown in the lower-right panel, while large part of Type-I parameter space (red) can be consistent, the entire parameter space for Type-F can be distinguishable from this benchmark point. The difference between the solid and dashed regions shows that the impact of the loop corrections is indeed large. 
 
In our analyses with the loop corrections, we have kept  $m_H=m_A=m_{H^\pm}=800$ GeV and $\sqrt{\lambda v^2}=300$ GeV to be the same for all the four types of 2HDMs.  Given that the loop corrections to the Higgs couplings depend on $\lambda v^2$ and $m_{H/A/H^\pm}$, allowing these parameters to vary in addition to $\cos(\beta-\alpha)$ and $\tan\beta$ will inevitably increase the 95\% C.L.  contour region, making it potentially more ambiguous in differentiating different types of 2HDMs.
 
\clearpage
\begin{figure}[!htbp]
\centering
\includegraphics[width=0.45\textwidth]{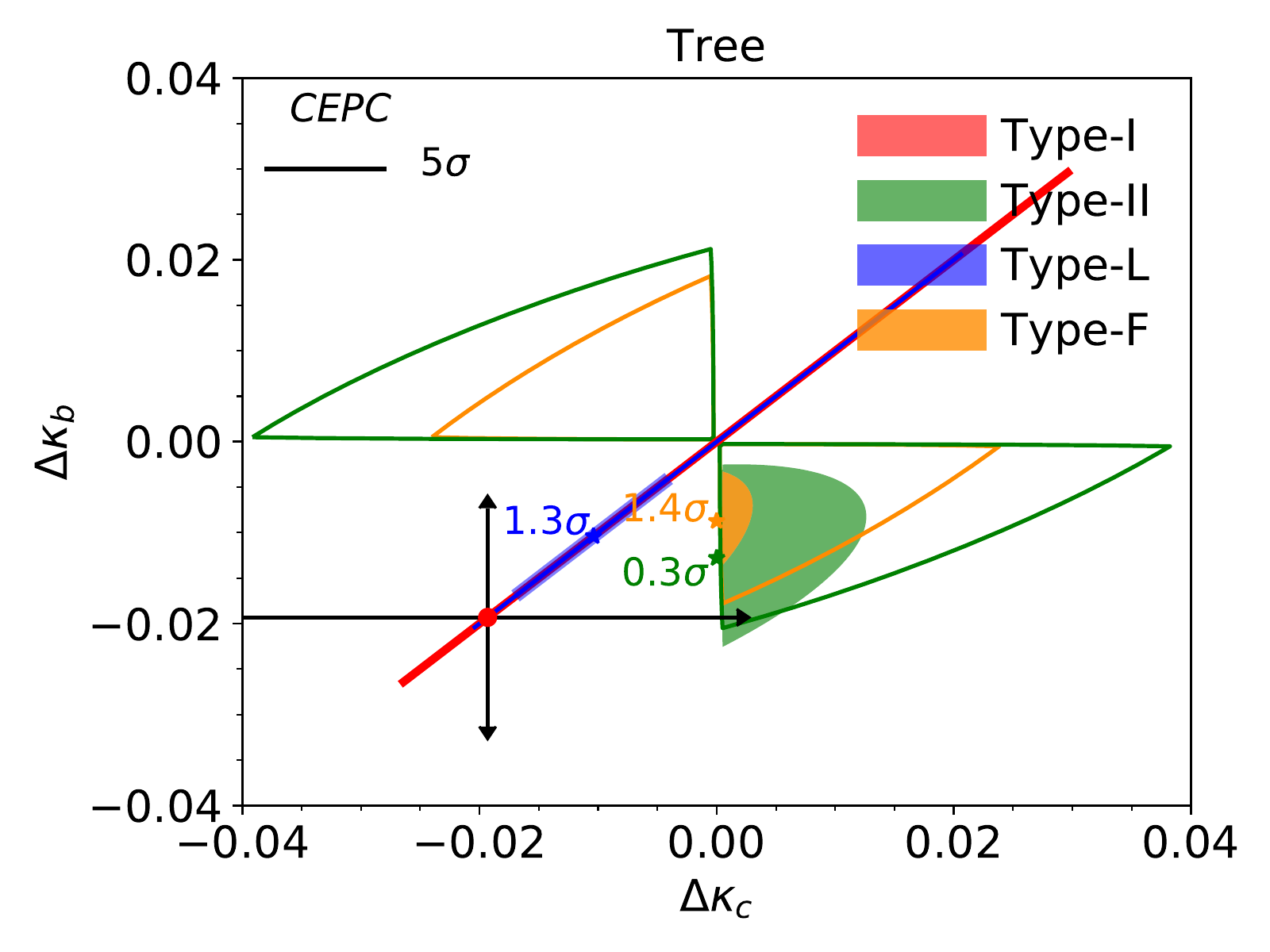}
\includegraphics[width=0.45\textwidth]{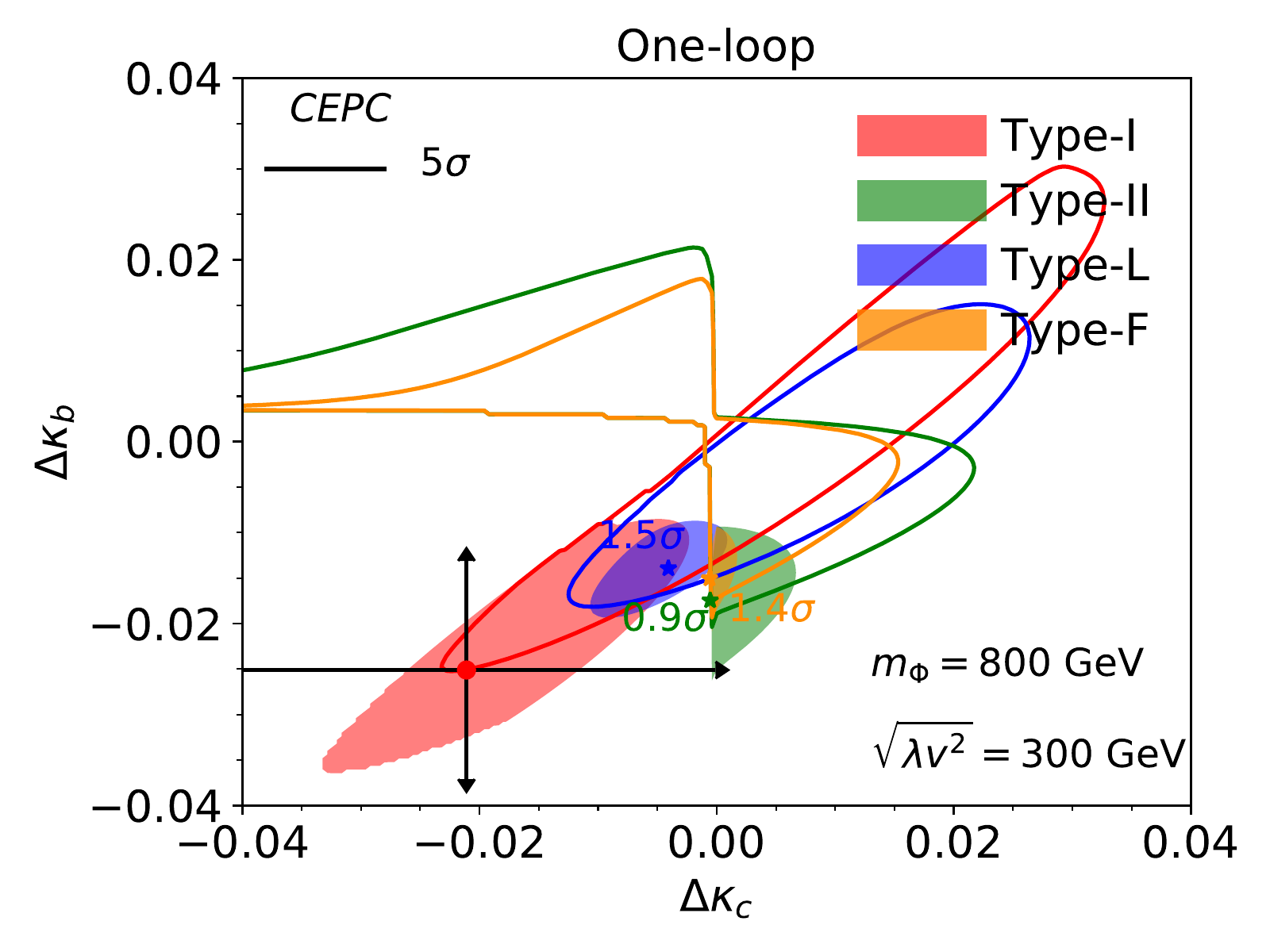}
\includegraphics[width=0.45\textwidth]{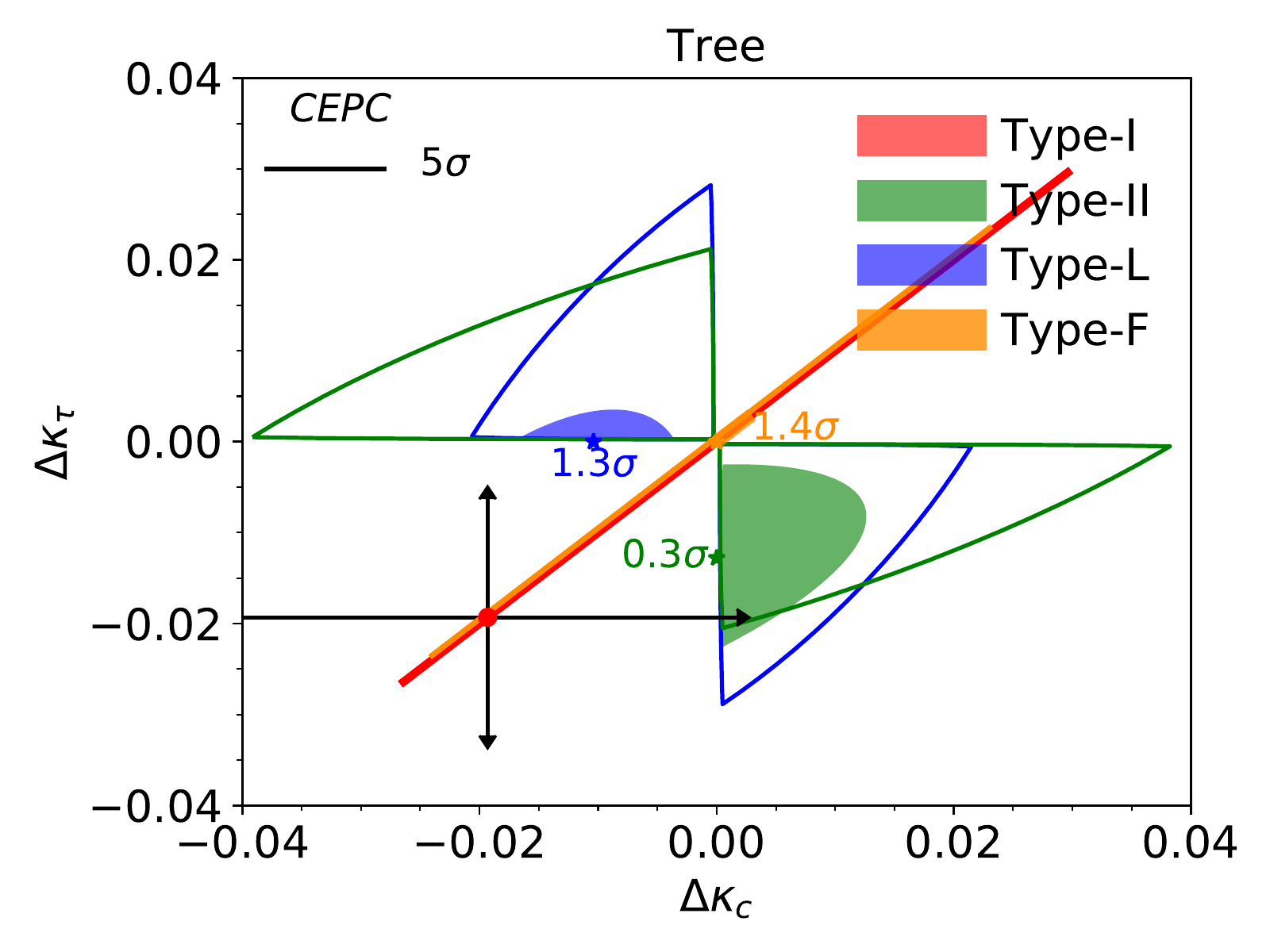}
\includegraphics[width=0.45\textwidth]{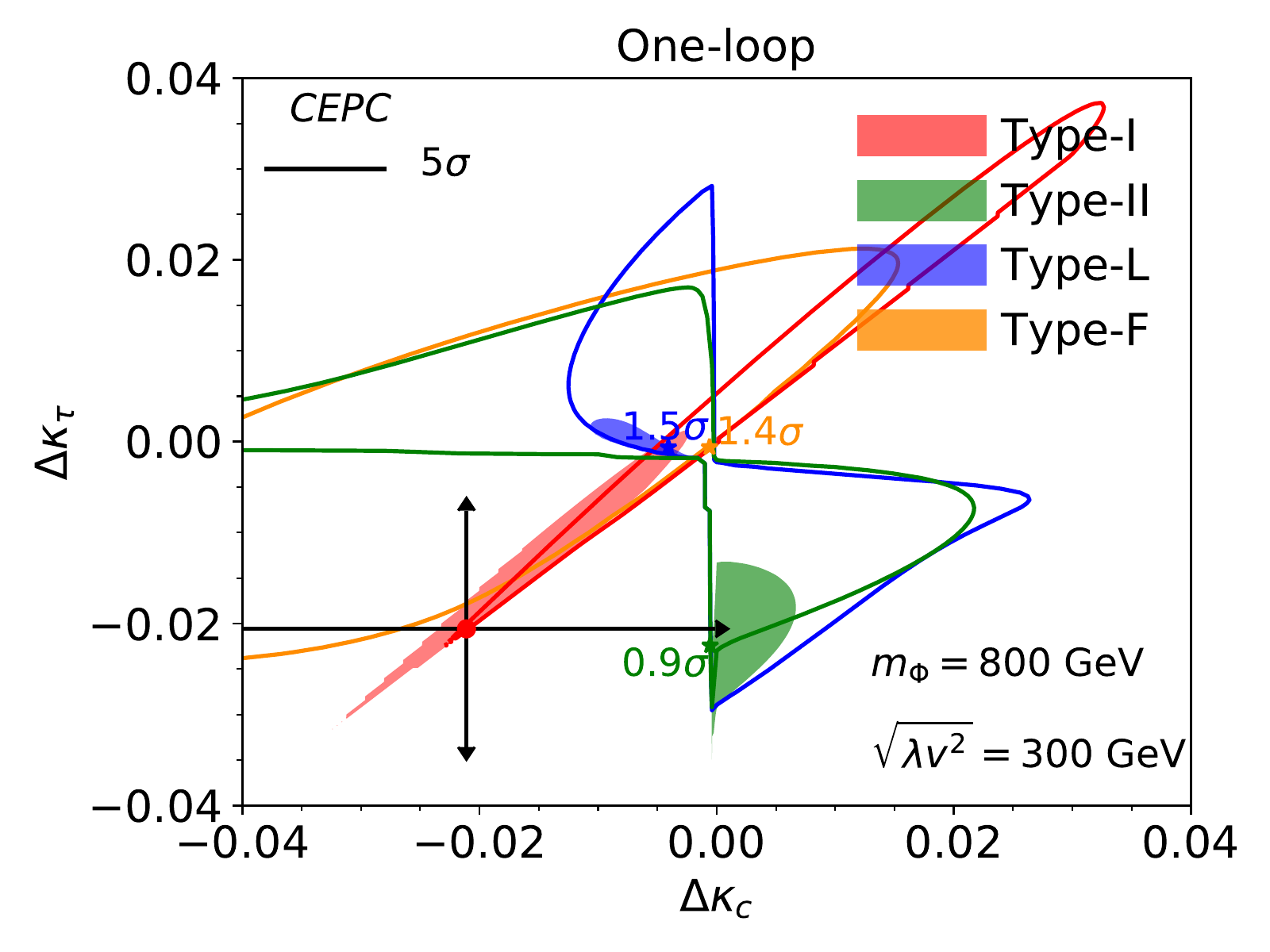}
\includegraphics[width=0.45\textwidth]{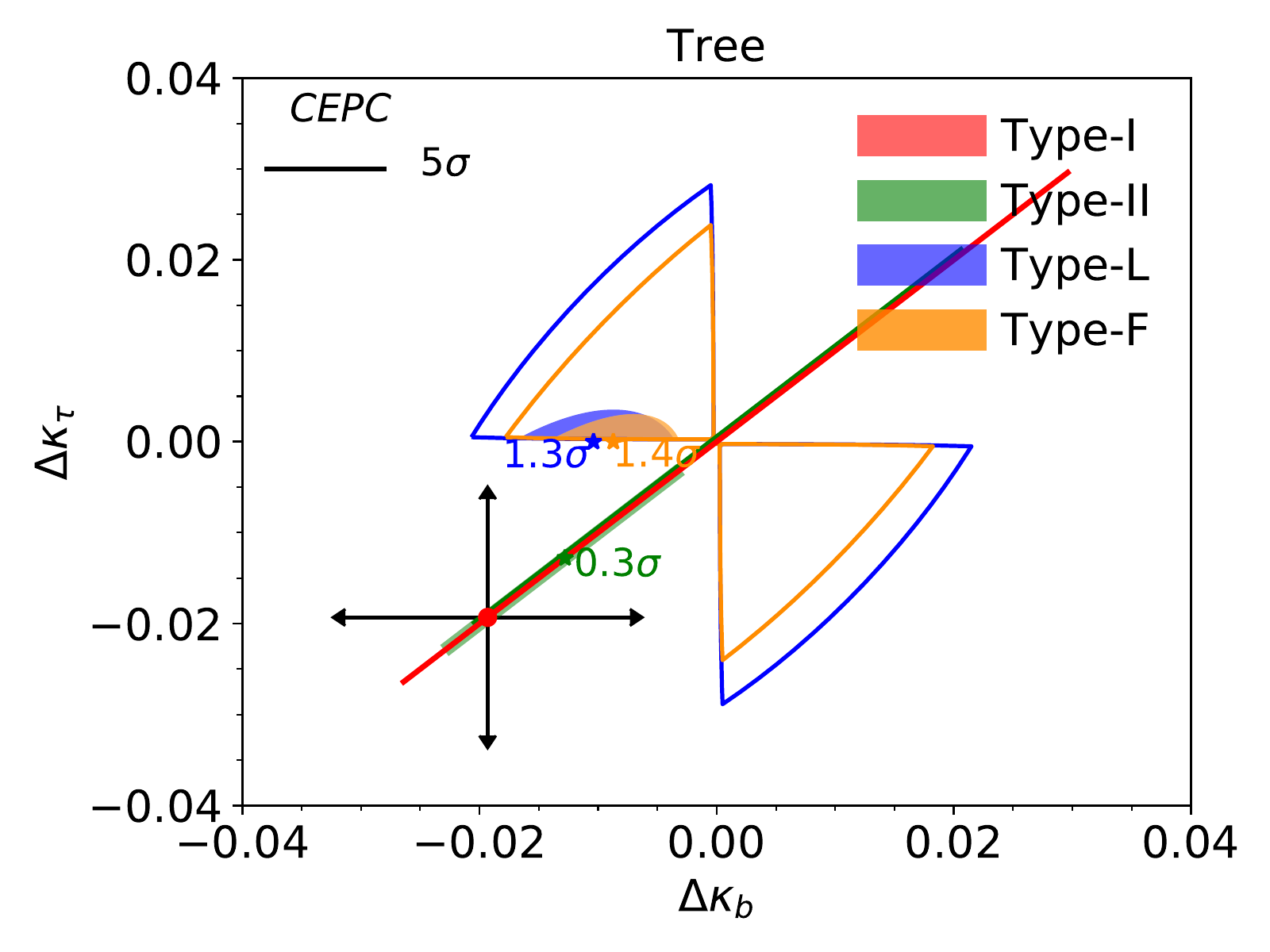}
\includegraphics[width=0.45\textwidth]{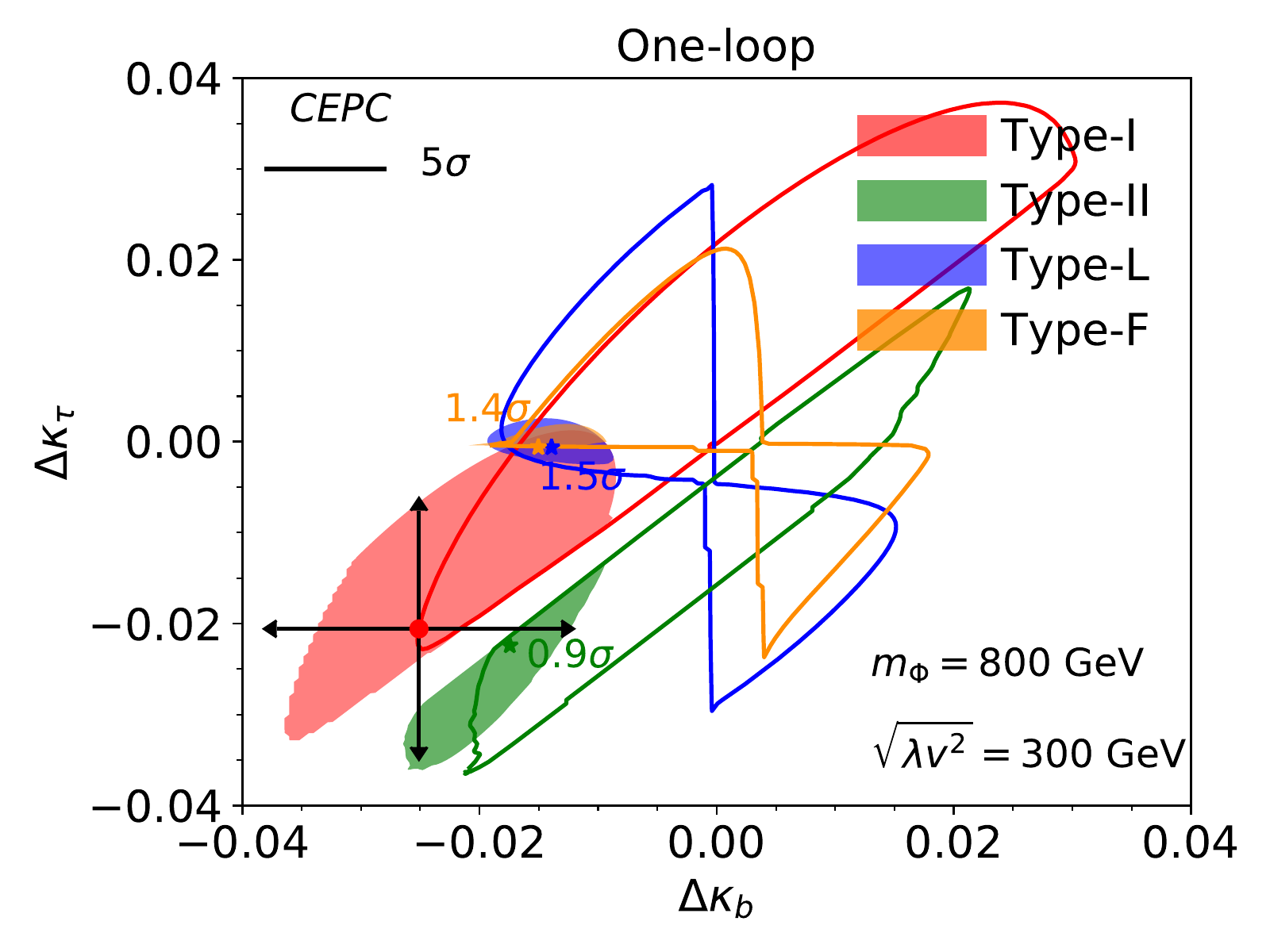}
\includegraphics[width=0.45\textwidth]{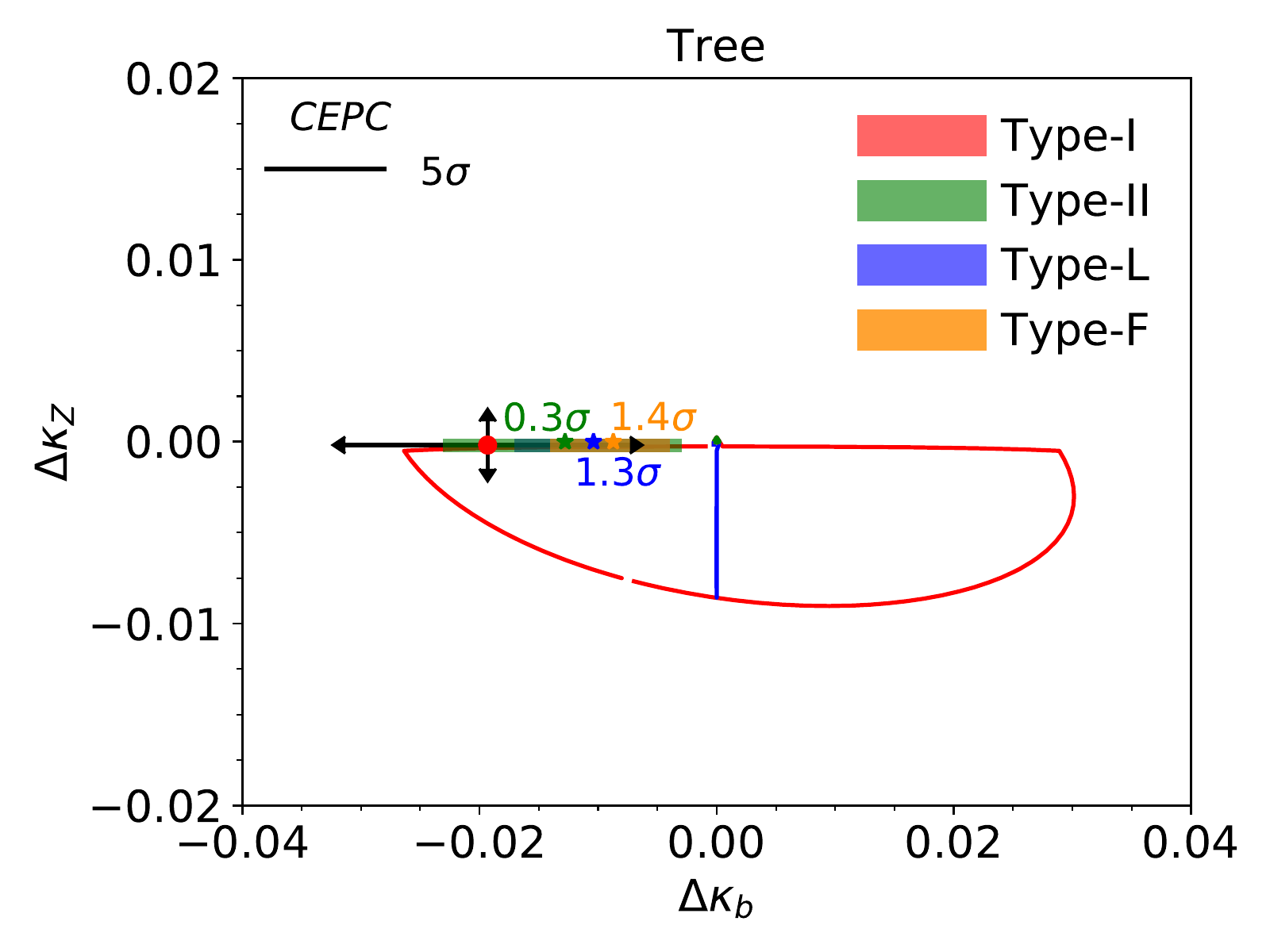}
\includegraphics[width=0.45\textwidth]{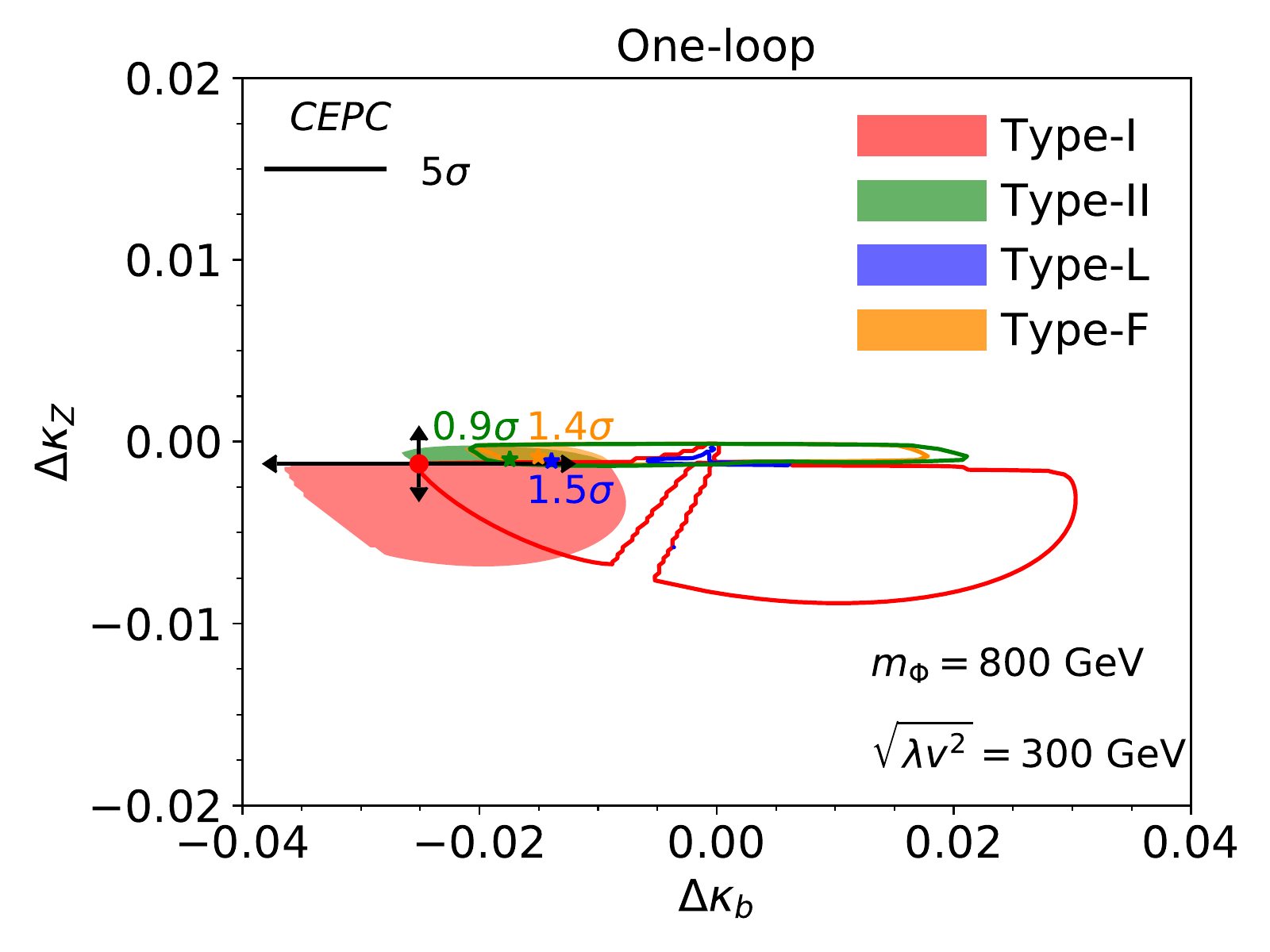}
\caption{Comparative study of the Benchmark point IA in Type-I with other types in $\kappa$ space. Shaded regions show the 95\% C.L.  discrimination regions, while regions enclosed by the solid curve are the CEPC $5\sigma$ discovery regions. The left (right) panels are for the tree-level (loop) results. Red, green, blue and orange colors refer to Type-I, II, L and F, respectively.
Stars of consistent colors mark the best fitting point, of which the significance of inconsistency are labelled beside. } 
\label{fig:typeIloop_smalltanb}
\end{figure}
\clearpage

To directly connect the theory parameters to the experimental observables, we take the small $\tanb$ Benchmark point IA in Type-I, marked by the red dot in~\autoref{fig:typeIloop_smalltanb}, with the crosses on top of the dot indicating the precision associated with each $\kappa_i$ at the CEPC. We present the regions that are indistinguishable from the other three types of 2HDMs in $\kappa$ space at 95\% C.L. as shaded areas in~\autoref{fig:typeIloop_smalltanb}. The left panels are for the tree-level results while the right panels include the one-loop results. Red, green, blue and orange colors refer to the points in Type-I, II, L and F, respectively. The shaded orange region in the right panel of the second row is too thin to be visible.   The best fitting points are marked by stars, with its significance labeled as well.    In each panel, we also show the $5\sigma$ CEPC discovery regions by the solid contour lines.  Note that the small gap in the red contour line of the bottom-left panel, as well as the deep indent in the bottom-right panel, is  caused by our scanning region of $\tanb\in [0.1,50]$. 

Comparing the solid $5\sigma$ discovery contours with the CEPC measurement in the left (tree level) and the right (loop level) panels, the $5\sigma$ discovery regions for Type-II (green) and F (orange) in the top two right panels exhibit a significant spread towards negative $\Delta\kappa_c$ region compared to their tree level counterparts. This is associated with the large stretch of the corresponding solid contour lines in \autoref{fig:loopvstree5sigma} towards very low $\tan\beta$ region: small $\tan\beta$ enhances the Yukawa couplings induced by the second Higgs doublet $\Phi_2$ as well as the loop corrections.

Given that Benchmark IA   corresponding to $\Delta \kappa_b=-0.025$, $\Delta\kappa_\tau=-0.021$ , $\Delta \kappa_c=-0.021$, $\Delta \kappa_Z=-0.001$ at one-loop level,   and the corresponding precisions of the experimental measurements of those Higgs couplings, there are small shaded regions in all three other types of 2HDMs that can not be distinguished  from this Type-I benchmark point at 95\% C.L.   However, most of the parameter space of the other three types of 2HDMs can be separable from the Benchmark IA.  In particular, given that most of the shaded region is inside the 5$\sigma$ discovery contour of the other three types of 2HDMs, except for $\kappa_b$ in the Type-II 2HDM, if a 5$\sigma$ discovery is made in any of these three 2HDMs outside the shaded region, it can be separable from Benchmark IA at more than 95\% C.L.

In $\kappa$-space, while the 95\% C.L. shaded regions tend to approach the benchmark point as much as they can, the discovery contours set a limitation on how much they can achieve. As a result, all best fitting points stop at the edge of the  discovery contour lines. So we can use the contour lines as guidance to understand how a benchmark point could be distinguished from other types. For the Benchmark IA marked on the panels, all other types show shaded regions from which it cannot be separated, but it is particularly difficult to distinguish it from the best fitting point in Type-II. This is because it has the green contour line of Type-II in its immediate neighborhood in the plane of $\Delta\kappa_\tau$ and $\Delta\kappa_b$, both of which are precisely measured in CEPC.   

In addition, going from Benchmark IA to IB in the right panels, one could move the red dot along the red contour line clockwise in the top panel and  counter-clockwise in the following three panels, given the dependence of $\Delta\kappa_i$ on $\tan\beta$ and $\cos(\beta-\alpha)$.   Although it seems harder to distinguish the red dot with other types as it gradually approaches the origin, the substantial deviation in $\kappa_Z$ in Type-I, as shown in the bottom panels,  makes it relatively easy to separate Benchmark IB from the other three types. This is also  reflected in \autoref{fig:compareloop}, in which Benchmark IB can be  distinguished  from all other types except for a tiny  region in Type-II.   

\section{Summary and Conclusions}
\label{sec:conclude}

The extension of the SM Higgs sector to two scalar doublets is well motivated in some theoretical scenarios for BSM physics. The additional Higgs bosons are thus amongst the most-wanted targets for searches for new physics in the current and future high-energy experiments. 

We considered four types of 2HDMs with different patterns of Yukawa couplings and a $\mathbb{Z}_2$ discrete symmetry, named Type-I, Type-II, Type-F and Type-L, outlined in Sec.~2. We laid out our analysis strategy and our fitting method in Sec.~3. In Sec.~4, we first presented the 95\% C.L.~allowed region under the current LHC Run-II limit in the $\cosba$-$\tanb$ plane, as shown in  \autoref{fig:lhcrun2limit}. The effects of loop corrections are generally small, except for large $\tan\beta$ region of the Type-I 2HDM when the loop effects are manifest,  giving a small tree-level distortion. Roughly speaking, the largest allowed ranges for $\cos(\beta-\alpha)$ at the 95\% C.L. are
\begin{align}
\nonumber
\text{Type-I:}&\quad (\cos(\beta-\alpha),\ \tan\beta) \sim (\pm 0.3,\ 2) ; \\
\nonumber
\text{Type-L:}&\quad (\cos(\beta-\alpha),\ \tan\beta) \sim (\pm 0.2,\ 1); \\
\nonumber
\text{Type-II, F:}&\quad (\cos(\beta-\alpha),\ \tan\beta) \sim (\pm 0.08,\ 1).
\end{align}

We further examined the $5\sigma$ discovery potential from the precision measurements for the four types of 2HDMs, and we found that most of the currently allowed region permits a discovery  at the CEPC and HL-LHC, as shown in  \autoref{fig:lhcrun2limitkappa} and \autoref{fig: lhcrun2limitkappaloop_300}, and  summarized in \autoref{fig:loopvstree5sigma}.   With the CEPC Higgs precision measurements,  the 5$\sigma$ discovery regions are 
\begin{align}
\nonumber
\text{Type-I:}&\quad \cos(\beta-\alpha)\lesssim -0.1\  {\rm or}\  \gtrsim 0.08\ {\rm for}\ 2\lesssim \tan\beta \lesssim 5; \\
\nonumber
\text{Type-L, II, F:}&\quad |\cos(\beta-\alpha)| \gtrsim 0.02\ {\rm for}\ \tan\beta \sim 1.
\end{align}
At small and large values of $\tan\beta$, the regions in $\cos(\beta-\alpha)$ are further tightened.

By studying the deviations in various couplings normalized to the SM value $\Delta\kappa_i\equiv\kappa_i-1$, for $i=t,b,\tau$ or $Z$, predicted by different types of 2HDMs, we presented their  correlation  in six  $\Delta\kappa_i$-$\Delta\kappa_j$ planes, as shown in  \autoref{fig:lhcrun2limitkappa} (tree-level) and \autoref{fig: lhcrun2limitkappaloop_300} (one-loop). Combining all six panels, we found that the four types of 2HDMs exhibit very distinct distributions in $\kappa$-space except for the common intersection at origin. In particular, measuring $\Delta\kappa_Z$ provides an immediate separation between Type-I and the other three types, if a significant deviation from the SM is observed. While the loop corrections weaken the capability to distinguish different types of 2HDMs, in particular due to the spread of the diagonal regions in Yukawa coupling correlations, combining all six panels still demonstrates the advantage in the model discrimination.  

Because of the characteristic features of the different 2HDMs, 
We further demonstrated the extent to which that they could be distinguishable from each other, if a discovery is made. We chose four benchmark points for illustration.  We found that a large part of the parameter space of the other types of 2HDMs can be distinguished from the benchmark point of the target model, as shown in \autoref{fig:compareloop}.  Including loop effects also shifts  the tree level results significantly. Our analyses has chosen a fixed value of the degenerate BSM Higgs mass $m_\Phi$ and soft $\mathbb{Z}_2$ breaking parameter $m_{12}^2$, for the sake of illustration.  
We further presented the discrimination ability of the 2HDMs in terms of the experimental measurements of $\kappa_i$ in \autoref{fig:typeIloop_smalltanb} for the small $\tanb$ benchmark point IA in Type-I. We see that once a 5$\sigma$ discovery is made in a particular type of 2HDM, it is very likely to be distinguished from other types of 2HDMs.

In summary, our analyses demonstrated the impressive potential for the future precision measurements on the SM Higgs precision observables at the Higgs factory, that could help to discover the BSM Higgs physics and to discriminate among the different incarnations of the two Higgs doublet models. 

\begin{acknowledgments}


We would like to thank S.~Heinemeyer for stimulating discussions. 
TH is supported in part by the U.S.~Department of Energy under grant No.~DE-FG02-95ER40896 and by the PITT PACC. SL and SS is supported  by the Department of Energy under Grant No.~DE-FG02-13ER41976/DE-SC0009913.
WS were supported by  the  Australian  Research  Council  Discovery  Project DP180102209. YW is supported by the Natural Sciences and Engineering Research Council of Canada (NSERC).  
\end{acknowledgments}

\bibliographystyle{JHEP}
\bibliography{references}

\end{document}